\documentclass{llncs}
\usepackage[latin1]{inputenc}
\usepackage{amsmath}
\usepackage{amsfonts}
\usepackage{amssymb}
\usepackage{graphics}
\usepackage{graphicx}
\usepackage{color}
\title{NLHB : A Non-Linear Hopper Blum Protocol}
\makeatletter
\let\@copyrightspace\relax
\makeatother

\makeatletter
\renewcommand\subsubsection{\@startsection{subsubsection}{3}{1mm}
{-\baselineskip}
{1\baselineskip}
{\normalfont\normalsize\bfseries}}
\makeatother
\newlength{\thirdtw}
\newlength{\fourthtw}
\newcommand{\medrarrow}{\overrightarrow{\hspace{\fourthtw}}}
\newcommand{\medlarrow}{\overleftarrow{\hspace{\fourthtw}}}
\author{Mukundan Madhavan\inst{1} \and Andrew Thangaraj\inst{1} \and Yogesh Sankarasubramaniam\inst{2} \and Kapali Viswanathan \inst{2}}
\institute{Indian Institute of Technology, Madras \and HP Labs India, Bangalore}
\begin{document}
\maketitle
\begin{abstract}
In this paper, we propose a light-weight provably-secure authentication protocol called the NLHB protocol, which is a variant of the HB protocol~\cite{hopperblum}. The HB protocol uses the complexity of decoding linear codes for security against passive attacks. In contrast, security for the NLHB protocol is proved by reducing passive attacks to the problem of decoding a class of non-linear codes\footnote{See Section~\ref{nlcodes} for the definition of the class of non-linear codes.} that are provably hard. We demonstrate that the existing passive attacks(\cite{fouque},\cite{imai}) on the HB protocol family, which have contributed to considerable reduction in its effective key-size, are ineffective against the NLHB protocol. From the evidence, we conclude that smaller-key sizes are sufficient for the NLHB protocol to achieve the same level of passive attack security as the HB Protocol. Further, for this choice of parameters, we provide an implementation instance for the NLHB protocol for which the Prover/Verifier complexity is lower than the HB protocol, enabling authentication on very low-cost devices like RFID tags. Finally, in the spirit of the HB$^{+}$ protocol, we extend the NLHB protocol to the NLHB$^{+}$ protocol and prove security against the class of active attacks defined in the DET Model. 

\end{abstract}
\section{Introduction}
The HB protocol was proposed in~\cite{hopperblum} as a low-complexity authentication algorithm that can be computed by human users. Its security is based upon the hardness of the ``Learning Parity in Noise'' (LPN) problem~\cite{lpnishard}, which is known to be NP-Hard. Though the protocol is secure against passive attacks, where the attacker is allowed only to eavesdrop on protocol communications, it was found to be vulnerable to active attacks, where the attacker could send spurious messages to the protocol participants. Having discovered this efficient active attack against the protocol, Juels and Weis~\cite{juelsandweis} proposed the HB$^{+}$ protocol as an alternative that could resist active attacks. The added complexity of the HB$^{+}$ protocol and the protocol's need for generation of many random numbers by the Prover rendered it more suitable for low-complexity RFID tags rather than human users.

Cryptanalysis of the HB authentication protocol has resulted in efficient solutions to the LPN problem. Notably, Levieil and Fouque~\cite{fouque} proposed the LF2 algorithm, which is an improved form of the BKW algorithm~\cite{bkw} for solving the LPN problem. Later, Carrijo \textit{et al.}~\cite{imai} proposed a probabilistic passive attack against HB and HB$^{+}$ protocols. These new solutions have significantly reduced the effective key-size of the HB protocol family that depend on the hardness of decoding linear codes for security against passive adversaries.

In this paper, we define and consider the UNLD problem, which is a decoding problem for a specific class of non-linear codes. We prove hardness of UNLD by reducing the LPN problem to the UNLD problem. Following this, we propose the NLHB protocol, which is a carefully constructed variant of the HB protocol. Security of NLHB against passive attacks is proved by reduction from UNLD to the passive attack problem. 

The basic idea behind the NLHB protocol is the use of a carefully-chosen non-linear Boolean function on the linear parities generated in the HB protocol. The use of this non-linear function does not affect the provable security of NLHB as a reduction from the provably-hard UNLD problem still works under a simple uniformity condition satisfied by the function. On the practical side, the use of the non-linear function considerably weakens the effectiveness of passive attacks like LF2 \cite{fouque} that depend on the linearity of the parities. Therefore, key efficiency is higher in NLHB when compared to HB.

For implementation, we demonstrate a certain quadratic form chosen from the general family of functions that we propose for the NLHB, which presents a specific low-cost candidate for the protocol. Using this candidate function, the complexity of the NLHB protocol is low enough that it can be implemented in low-cost devices such as RFIDs. Finally, we show that the Prover/Verifier complexity of NLHB protocol can be lower than that of the HB protocol because of the use of smaller keys.

Active attacks similar to those on the HB protocol are possible on the basic NLHB protocol. We demonstrate that the basic NLHB protocol can be extended to an NLHB$^+$ protocol, in the spirit of HB$^+$, for security in some active attack models. We show that the reductions for the HB$^{+}$ protocol as shown in~\cite{katzandshin,katzandsmith} work for the NLHB$^+$ protocol as well.

In summary, the main contribution of this paper is a low-cost, provably-secure extension of the HB protocol through the use of simple non-linear functions on parities. Because of the non-linearity, the proposed NLHB protocol has better resistance to known passive attacks on the HB family resulting in higher key efficiency and cheaper implementations. Also, the NLHB protocol can be modified in the spirit of the several known modifications of the HB protocol to obtain better security against different classes of active attacks.  

The paper is organized as follows. In Section 2, we give a brief introduction on the HB and HB$^{+}$ protocols, related security models and the ``Learning Parity in Noise" (LPN) problem. In Section 3, we describe the UNLD problem, a type of non-linear code decoding problem and prove its NP-Hardness. This is followed by a description of the NLHB protocol and its security proofs. Section 4 contains discussions on the resistance of the protocol to passive attacks and its Prover complexity. This is followed by the proposition of the NLHB$^{+}$ protocol and its security proofs in Section 5. Section 6 concludes the paper.
\section{The HB And HB$^{+}$ Protocols} 
\subsection{HB Protocol} 
\begin{figure}[!htbp]
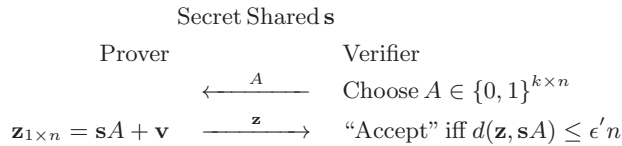

\begin{eqnarray*}
& \mathrm{Secret} \, \mathrm{Shared} \, \mathbf{s} & \\
\mbox{Prover} && \mbox{Verifier} \\
 &\stackrel{A}{\medlarrow}&   \mathrm{Choose} \, A \in \{ 0,1 \} ^{k \times n} \\
\mathbf{z}_{1 \times n}=\mathbf{s}A + \mathbf{v} & \stackrel{\mathbf{z}}{\medrarrow}& \mathrm{``Accept"} \, \mathrm{iff} \, d(\mathbf{z}, \mathbf{s}A) \leq \epsilon' n
\end{eqnarray*}
\caption{Parallelized version of the HB protocol}
\label{fighb}
\end{figure}

\hspace*{9mm}The HB protocol is a symmetric-key authentication protocol. The Prover and Verifier share a random $k$-bit secret key $\mathbf{s}$~\footnote{Refer Appendix D for notations}. The protocol has two public probability parameters $\epsilon, \epsilon'\in \mbox{ } ]0,\frac{1}{2} [ \mbox{ }$ such that $\epsilon < \epsilon'$. To authenticate, the Verifier sends a random $k$-bit challenge vector $\mathbf{a}$. The Prover, in turn, calculates the binary dot-product $\mathbf{s}.\mathbf{a}$ and replies to the Verifier with $z=\mathbf{s}.\mathbf{a} + v$, where $v$ is a Bernoulli random variable that takes the value 1 with probability  $\epsilon$ and $+$ denotes XOR addition. This process is repeated $n$ times. At the end of $n$ repetitions, the Verifier returns an ``Accept" message iff atmost $\epsilon'n$ responses are ``wrong", i.e, different from dot-products of the secret and the corresponding challenges. This process, which constitutes one authentication session can be parallelized as shown in Figure ~\ref{fighb}.\\
\begin{figure}[!htbp]
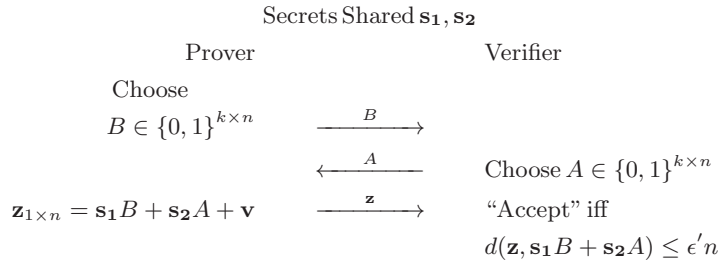

\begin{eqnarray*}
& \mathrm{Secrets} \, \mathrm{Shared} \, \mathbf{s_{1}},\mathbf{s_{2}} & \\
\mbox{Prover} && \mbox{Verifier} \\
\mathrm{Choose} \, \, \, \; \; \; \; \; \; \; \; &&\\
B\in \{ 0,1 \} ^{k \times n} & \stackrel{B}{\medrarrow} &\\
 &\stackrel{A}{\medlarrow}&   \mathrm{Choose} \, A \in \{ 0,1 \} ^{k \times n} \\
\mathbf{z}_{1 \times n}= \mathbf{s_{1}}B + \mathbf{s_{2}}A + \mathbf{v} & \stackrel{\mathbf{z}}{\medrarrow}& \mathrm{``Accept"} \, \mathrm{iff}\\
&&d(\mathbf{z}, \mathbf{s_{1}}B + \mathbf{s_{2}}A) \leq \epsilon' n
\end{eqnarray*}
\caption{Parallelized version of the HB$^{+}$ protocol}
\label{fighb+}
\end{figure}
\hspace*{9mm} In the parallelized form, the Verifier challenges the Prover with a random $k \times n$ matrix, to which the Prover responds with $\mathbf{z}=\mathbf{s}A + \mathbf{v}$. Here, the bits of the vector $\mathbf{v}$ are all i.i.d Bernoulli random variables with parameter $\epsilon$ and the multiplication between the vector $\mathbf{s}$ and $A$ is over the binary field $GF(2)$. The response vector $\mathbf{z}$ is a $n$-bit vector and the Verifier responds with ``Accept" iff $d(\mathbf{z},\mathbf{s}A) \leq \epsilon' n$, where $d(.)$ denotes Hamming distance. The parameters $\epsilon$,$\epsilon'$, and $n$ are fixed so that both the probability of rejecting an honest Prover as well as the probability of positively authenticating an attacker giving random responses are negligible (\cite{fouque}, Figure 2).
 The HB Protocol has been proven secure in the Passive attack model as defined below.
\begin{definition}[Passive attack model (\cite{juelsandweis},\cite{katzandshin})]
In this model, the adversary algorithm is two-phased. In the first phase (called the query phase), the adversary has access to the transcripts from an arbitrary number of authentication sessions between an honest Prover and Verifier. In the second phase (called the cloning phase), the adversary tries to impersonate an honest Prover to the Verifier.
\end{definition}
However, the HB protocol is not secure against active attacks \cite{juelsandweis}.
\subsection{HB$^{+}$ Protocol}
\hspace*{9mm}The HB protocol is susceptible to a simple active attack. In this attack, the attacker repeatedly challenges an honest Prover with the same challenge, and by majority vote over these multiple responses, decides (with high confidence) on the noise-free response. This is repeated for $k$ linearly independent challenges, following which, the secret key is easily found using a Gaussian elimination over the system of linear equations defined by these $k$ challenge-response pairs \cite{juelsandweis}. Thus, the active attacker need not solve the LPN problem to attack the HB protocol.\\
\hspace*{9mm}To counter such attacks, Juels and Weis \cite{juelsandweis} proposed the HB$^{+}$ protocol (Figure~\ref{fighb+}). Instead of a single secret, the Prover and Verifier share two $k$-bit secret keys $\mathbf{s_{1}}$ and $\mathbf{s_{2}}$~\footnote{The sizes of these secret can be different. This paper shall consider them to be of same size without loss of generality.}. In its parallel form, the HB$^{+}$ protocol can be described as follows. The Prover starts an authentication session by sending a random ``blinding" matrix $B$ to the Verifier, which in turn replies with its own random challenge-matrix $A$. On receiving $A$, the Prover responds with $\mathbf{z}=\mathbf{s_{1}}B + \mathbf{s_{2}}A + \mathbf{v}$. Here, $A$ and $B$ are $k \times n$ matrices, and $\mathbf{v}$ has the same definitions as in the HB protocol. The Verifier responds with an ``Accept" decision iff $d(\mathbf{z}, \mathbf{s_{1}}B+ \mathbf{s_{2}}A) \leq \epsilon'n$. Now, when an active attack is mounted, the attacker still has to solve an LPN instance on the matrix $B$.\\
\hspace*{9mm} The HB$^+$ protocol is secure against both passive attacks as well as active attacks in a model known as the ``DET" attack model.
\begin{definition}[DET Attack Model(\cite{juelsandweis},\cite{katzandsmith})] In this model, attacks are two-phased. In the first(query) phase, the adversary can interact with an honest Prover an arbitrary number of times. In second (cloning) phase, the adversary interacts with the Verifier and attempts impersonation.
\end{definition}
\textit{Significance Of The ``DET" Model: }Juels and Weis discuss the significance of the ``DET" model in \cite{juelsandweis}, [Appendix A]. Even though this model does not include Man-In-Middle attackers and does not give an attacker access to Verifier decisions at the end of authentication, it is an important security model in a context where the adversary has to forge a valid Prover without the attack being detected. Since attacks in the more powerful prevention-based models like GRS-MIM \cite{GRS} may not be undetected attacks, the prevention-based model is not the ideal model in all scenarios. As an example, in a setting where the Verifier would report repeated authentication failures from a Prover, the detection-based model is more suitable. 
\subsection{The LPN Problem and Passive Attacks}
\label{lf2explained}
\begin{definition}[LPN Problem \cite{juelsandweis}] Let $\mathbf{s}$ be a random binary $k$-bit vector. Let $\epsilon \in]0,\frac{1}{2}[$ be a constant error parameter. Let $A$ be a random $k \times n$ matrix, and let $\mathbf{v}$ be a random $n$-bit vector such that $\mathrm{wt}(\mathbf{v})\leq  \epsilon n$, where $\mathrm{wt}(\mathbf{v})$ denotes the Hamming weight of $\mathbf{v}$. Given $A$, $\epsilon$ and $\mathbf{z}=(\mathbf{s}A)+ v$, find a $k$-bit vector $\mathbf{s'}$ such that $d(\mathbf{z},\mathbf{s'}A)\leq \epsilon n$.
\end{definition}
For large $n$, this is equivalent to finding the vector $\mathbf{s}$. The LPN problem has been proven to be NP-Hard \cite{lpnishard} and is conjectured to be average-case hard \cite{hopperblum}. The BKW algorithm, which was the best-known algorithm to solve the LPN problem when the HB protocol was proposed, has a high complexity and requires a large number of challenge-response pairs $\langle A,\mathbf{z} \rangle$ to obtain a solution. The LF2 algorithm \cite{fouque}, which is an improvement over the BKW algorithm, has considerably lesser complexity and needs lesser challenge-response pairs for its solution. Later, a probabilistic attack on the LPN problem was proposed by Carrijo \textit{et al.} in \cite{imai}. These attacks have reduced the effective key-size of the HB protocol, necessitating higher key-sizes. We first describe the LF2 attack, followed by the attack proposed by Carrijo \textit{et al}.\\
\hspace*{9mm}When the key-size is high, exhaustive search over the space of all possible keys is intractable. So, the LF2 attack aims to estimate few bits of the key at a time. The attack involves adding the columns of the challenge-matrix $A$ (and the corresponding responses) so that only the first few (say $b$) rows have non-zero entries in the resulting matrix. This addition causes two different changes. First, adding two or more noisy responses results in an increased chance of the new response being wrong. So, the apparent Bernoulli parameter in this new set of equations is higher. However, the second and more important change is that, since only the secret bits corresponding to the $b$ non-zero rows play a role in the multiplication $\mathbf{s}A$, the attacker can now find these $b$ bits in isolation by running an exhaustive search over $2^{b}$ possibilities. So, by running an exhaustive search over a space of size $2^{b}$ (which is much smaller than $2^{k}$), the first $b$ bits of the original key can be found. Repeating this process for the second $b$ rows, and so on, gives the whole key. Thus, the attack depends heavily on the fact that the Prover's response is a noisy version of some codeword from the linear code having $A$ as its generator.\\
\hspace*{9mm}A second new passive attack was also proposed by Carrijo \textit{et al.} \cite{imai}. This attack tries to pick noise-free bits from the response vector and find the key through Gaussian elimination on the system of equations formed from these bits alone. So, this attack too, depends on the Prover's response being the noisy version of a codeword of the linear code generated by $A$.\\
\hspace*{9mm}As a consequence of these attacks, a LPN instance using as many as 512 bits of secret can be attacked with a complexity of just $2^{80}$ operations.

\section{The UNLD Problem and the NLHB Protocol}
\label{nlcodes}
The main idea in this paper is to replace the linear parity generation part $sA$ in the HB protocol with a non-linear version $f(sA)$ for a suitable public function $f$. The following characteristics are desirable for such a function $f$:
\begin{enumerate}
\item The function $f$, assumed to be public, must allow for the reduction of hardness from decoding problems to passive attacks on the protocol.
\item The function $f$ must be simple enough to implement on low-cost devices.
\item The function $f$ must provide better resistance to known passive attacks that solve the LPN problem.
\item The function $f$ should allow extensions such as HB$^+$ for security against active attacks. 
\end{enumerate}
We now describe a specific class of non-linear Boolean vector functions and discuss some of its properties that will be used in the security reductions. We discuss the other characteristics like implementation-cost and passive attack resistance in later sections.
\subsection{The Function $f$}
Let $D$ and $p$ be positive integers such that $D = n-p$ ($n$ is as described in the HB protocol). We propose the following construction for the NLHB protocol function $f$. Each bit $y_i; i \in [1,..,D=n-p]$ of the output $\mathbf{y}=f(\mathbf{x}); \mathbf{y} \in \{0,1\}^D, \mathbf{x} \in \{0,1\}^n$ will be computed as 
 \begin{equation}
 y_i = x_{i} + g([x_{i+1},..,x_{i+p}]),
 \label{generalf}
 \end{equation}
where $x_1,..,x_{n}$ are the bits of $\mathbf{x}$ and $g$ is a $p$-bit to 1-bit Boolean function containing strictly non-linear terms. Below, we list some important properties for this class of functions.
\begin{enumerate}
\item $f$ : $\{0,1\}^{n} \Rightarrow \{0,1\}^{D}$ 
\item $f$ is a non-linear function.
\item For uniformly distributed $\mathbf{x} \in \{0,1\}^{n}$, $f(\mathbf{x})$ is uniformly distributed in $\{0,1\}^{D}$.
\end{enumerate}
A proof of Property 3 is provided in Appendix \ref{proofofuniformity}. Intuitively, it can be said that the function $g([x_{i+1},..,x_{i+p}])$ causes the output bits $y_i$ to be non-linearly related to $\mathbf{x}$ and the component $x_i$ helps in balancing the output bit $y_i$. 

As a specific example, the function defined by
 \begin{equation}
 y_i = x_{i} + x_{i+1}x_{i+2} + x_{i+2}x_{i+3} + x_{i+3}x_{i+1}, 1\leq i\leq D
 \label{fcand}
 \end{equation}
is a part of this function family when using $p=3$. The uniform distribution property for $p=3$ can be readily verified by exhaustively determining the joint distribution of $\{y_i,y_{i+1},y_{i+2},y_{i+3}\}$ for a fixed $i$. When we set $p=3$, the function $f$ will take a $n$-bit vector $\mathbf{x}$ and map it onto a $D=(n-3)$ bit response vector. As we can see, members of this family like the one described in~\eqref{fcand} require very low additional complexity (only 3 AND gates and 3 XOR gates in this case)for implementation and their use in any protocol's implementation will add very little complexity. This can easily be accomodated into any RFID tag, however cheap. 

In the next section, we describe how this function family can be used to create a robust protocol. In later sections, we use our specific candidate to demonstrate how their use in the protocol leads to increased passive attack resistance while still maintaining low implementation complexity. However, we would like to point out that our proofs of security hold for all functions in the general class of functions in \eqref{generalf}.
\subsection{UNLD Problem} 
\label{NLDDef}
Suppose $A_{k \times n}$ is the generator matrix of a linear code. Then all vectors of the form $\mathbf{s}A$ are codewords of this code. When we apply the function $f$ to these codewords $\mathbf{s}A$, i.e, we compute $f(\mathbf{s}A)$, the set of vectors $\{f(\mathbf{s_i}A)\}_{i=1}^{2^k}$ at the output can be viewed as a non-linear code. 

We now define the UNLD problem, which (in words) is the problem of decoding the class of non-linear codes defined by $f$ and $A$ as $\{f(\mathbf{s_i}A)\}_{i=1}^{2^k}$.
\begin{definition}[UNLD Problem] Let $\mathbf{s}$ be a random $k$-bit binary vector. Let $\epsilon \in ]0,\frac{1}{2}[$ be a constant error parameter. Let $A$ be a random $k \times n$ binary matrix and let $\mathbf{v}$ be a random $D$-bit vector such that $\mathrm{wt}(\mathbf{v}) \leq \epsilon D$, where $\mathrm{wt}(\mathbf{v})$ denotes the Hamming weight of $\mathbf{v}$. Given $A,\epsilon$ and $\mathbf{z}=f(\mathbf{s}A) + \mathbf{v}$, find the $k$-bit vector $\mathbf{s}$.
\end{definition}

We prove the hardness of the UNLD problem by reducing a random instance of the LPN problem, which is known to be NP-Hard to solve, to the UNLD problem. To show the reduction, we consider an existential algorithm $X$ that can solve the UNLD problem. We construct an algorithm $S$, which can solve a random LPN instance, when given access to $X$.

\begin{theorem}[LPN reduces to UNLD]
Let $A$ be a random $k \times n$ matrix, $\mathbf{v'}$ be a $(n-p)$-bit Bernoulli noise vector, and $\mathbf{s}$ be a random $k$-bit vector. Suppose there exists a probabilistic polynomial-time (PPT) algorithm $X$ with input $\langle A, \mathbf{y}=f(\mathbf{s}A)+\mathbf{v'} \rangle$ that can output $\mathbf{s}$ with probability atleast $\delta$. Then, there also exists a PPT algorithm $S$ that can solve a LPN problem instance $\langle G_{k \times n'},\mathbf{z}=\mathbf{m}G + \mathbf{v} \rangle$ for randomly chosen $\mathbf{m}$, Bernoulli noise vector $\mathbf{v}$ and $n' \leq \frac{(n-1)}{p} , k < n'$ with probability at least $\delta$.
\label{thm1}
\label{lpntounld}
\end{theorem}
\begin{proof}
Let $\mathbf{z} = [z_1,...,z_{n'}]$ and $\mathbf{v} = [v_1,...,v_{n'}]$ be the constituent bits of the vectors described above. The algorithm $S$, having access to algorithm $X$ works as follows to solve a random LPN instance $\langle G,\mathbf{z} \rangle$ passed to it.
\begin{enumerate}
\item Pick $r_i$ for $1 \leq i \leq n'-1$ such that  $r_i \geq (p-1), \sum_{i=1}^{n'-1} r_i = n-p-n'$.
\item Insert $r_i$ Bernoulli bits between bit $z_i$ and $z_{i+1}$ of $\mathbf{z}$ for $1 \leq i \leq n'-1$. This gives rise to the vector $\mathbf{y}_{(n-p)}=[z_1, \, b_1 \, b_2 \, ... \, b_{r_1},\, z_2, \, b_{r_1 +1} \, ... \, b_{r_1+r_2},\, z_3,\, .....b_{n-p-n'},\, z_{n'}]$. 
\item Insert $r_i$ columns of zeros in between columns $i$ and $i+1$ of $G$ ($1 \leq i \leq n'-1$) to get the matrix $A$. Insert $p$ columns of zeros after the last column of $A$. Now, the dimension of $A$ is $k \times n$ and $A$ is of the form $A = [\mathbf{g_1} \underline{0} \underline{0}..\underline{0} \mathbf{g_2} \underline{0} \underline{0}..\underline{0}..... \mathbf{g_{n'}}\underline{0}\underline{0}..\underline{0}]$, where $\mathbf{g_i}$ are the columns of $G$.
\item Pass $\langle A,\mathbf{y} \rangle$ to $X$ and get back $\mathbf{m'}$.
\item Return $\mathbf{m'}$ as the estimate of the LPN secret $\mathbf{m}$.
\end{enumerate}
We now show that $S$ succeeds with probability at least $\delta$. Consider the vector $\mathbf{\overline{x}}=\mathbf{m}A$. We can see that $\mathbf{\overline{x}}  = [x_1 00..0 x_2 00..0 x_3 00.0....x_{n'} 00..0]$, where $[x_1, x_2,...,x_{n'}]$ are the bits of $\mathbf{x}=\mathbf{m}G$. We also see that, since $g$ has only non-linear terms (i.e each term in $g$ is some kind of product of at least two input bits) and $r_i \geq (p-1)$, the vector $f(\mathbf{\overline{x}})$ can be written as $f(\mathbf{\overline{x}}) = [x_1 00..0 x_2 00..0 x_3.....00 x_{n'}]$, as all the product terms from $g$ go to zero.\\
  Let this new vector $f(\mathbf{\overline{x}})$ be called $\mathbf{x'}$. So, the vector $\mathbf{y}$ is of the form $\mathbf{x'} + \mathbf{v'}$ where \\ $\mathbf{v'} = [v_1, \, b_1 \,  b_2 \, ... \, b_{r_1},\, v_2,\, b_{r_1+1}\, ...\, b_{r_1+r_2},\, v_3,\, .....b_{n-p-n'},\, v_{n'}]$ Here, $v_i$ are the Bernoulli bits since they are part of the LPN noise vector $\mathbf{v}$ and $b_i$ are picked to be Bernoulli bits. So, $\mathbf{y} = f(\mathbf{m}A) + \mathbf{v'}$, where $\mathbf{v'}$ is a Bernoulli noise vector. Hence, by definition, $X$ will return $\mathbf{m}$ with probability at least $\delta$. Since $S$ succeeds whenever $X$ succeeds, the probability of success of $S$ is at least $\delta$. \qed
\end{proof}
We note that it is always possible to pick $r_i$ satisfying the condition in Step 1 for any $n$ and $n' \leq \frac{n-1}{p}$. As an example, one could initially fix $r_i = (p-1);\forall i$. Then $\sum r_i =(n'-1)(p-1) = n'p -p-n'+1$. Then one can add the difference $(n-p-n')-(n'p-p-n'+1) = n-n'p-1$ (which is always positive because of the upper bound on $n'$) to say, $r_1$, giving us a new set $\{r_i\}$ satisfying the conditions in Step 1 for any $n$.
\subsection{NLHB Protocol}
\begin{figure}[!htbp]
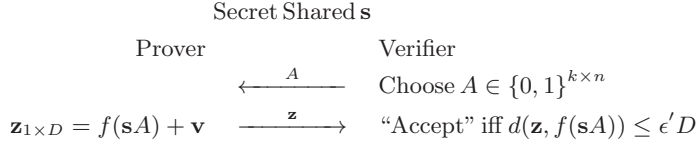

\begin{eqnarray*}
& \mathrm{Secret} \, \mathrm{Shared} \, \mathbf{s} & \\
\mbox{Prover} && \mbox{Verifier} \\
 &\stackrel{A}{\medlarrow}&   \mathrm{Choose} \, A \in \{ 0,1 \} ^{k \times n} \\
\mathbf{z}_{1 \times D}=f(\mathbf{s}A) + \mathbf{v} & \stackrel{\mathbf{z}}{\medrarrow}&\mathrm{``Accept"} \, \mathrm{iff} \, d(\mathbf{z}, f(\mathbf{s}A)) \leq \epsilon' D
\end{eqnarray*}
\caption{Parallelized version of the NLHB protocol}
\label{figNLHB}
\end{figure}

\hspace*{9mm}Having established the hardness of the UNLD problem, we now propose the NLHB protocol that is based on this problem. Figure \ref{figNLHB} shows one session of the NLHB protocol. The Prover and Verifier share a $k$-bit secret $\mathbf{s}$. The Verifier transmits a random $k \times n$ challenge matrix $A$ to the Prover. On receiving this, the Prover computes $f(\mathbf{s}A)$. Then, it computes $\mathbf{z}=f(\mathbf{s}A) + \mathbf{v}$, where $\mathbf{v}$ is a noise-vector whose bits are all independently distributed according to the Bernoulli distribution with parameter $\epsilon$, just like the noise vector in the HB protocol. Here $\mathbf{s}A$ is a $n$-bit vector and $\mathbf{z}$ is a $D$-bit vector. On receiving $\mathbf{z}$, the Verifier checks whether $d(\mathbf{z},f(\mathbf{s}A)) \leq \epsilon'D$. Iff this is true, it returns ``Accept". Here too, $\frac{1}{2} > \epsilon' > \epsilon$. Further since the noise-vector is of length $D$, $D$ has to be large enough ($\approx$1000) and ($D$,$\epsilon$,$\epsilon'$) have to satisfy the conditions satisfied by the HB protocol parameters ($n$,$\epsilon$,$\epsilon'$)(see Figure 2 of \cite{fouque}). For example, $D=1164$, $\epsilon=.25$ and $\epsilon'=.348$ is a possible parameter set.\\

\hspace*{9mm}Due to the non-linearity property of $f$, $f(\mathbf{s}A)$ is some unknown codeword of the random non-linear code $\{ f(\mathbf{s_i}A)\}_{i=1}^{2^k}$ and $\mathbf{z}$ is the noisy form of this codeword. To find the secret $\mathbf{s}$, the attacker now has to decode this random non-linear code instead of the linear code with generator matrix $A$. We will show in Section \ref{lf2attackrepulsed} that existing passive attacks on HB protocol family do not work on our protocol. Our proofs of security are valid for a general class of functions. However, in Section \ref{imploff}, we demonstrate that, for certain choices of $f$ within this family, the protocol complexity is very low.
\subsection{Security Proofs For NLHB In Passive Model}
The proof of security for NLHB in the passive model involves reductions from the UNLD problem to the forging of the NLHB protocol in the passive model. It is detailed in Theorems \ref{thm2} and \ref{thm3}. These theorems are broadly based on the proof of security given for the HB protocol in (\cite{katzandshin},\cite{katzandsmith}), with suitable modifications and additions to support the function $f$. Here, we first prove a technial lemma of our own, that is crucial to the proof of security. We then, use this lemma in the formal proof of our Theorem \ref{thm2}. Since the other parts of the proofs of Theorem \ref{thm2} and \ref{thm3} are similar to those in (\cite{katzandshin},\cite{katzandsmith}), we simply give a brief outline here, and delegate the formal proving to the Appendix. 

We first explain some brief notations needed for understanding the proof.\\
\newline
\textbf{Notations In The Proof:}
\begin{enumerate}
\item The distribution ${\cal A}_{\mathbf{s},\epsilon,f}$ is the distribution followed by the $(kn+D)$-length bitstrings in the transcript of one authentication session of the NLHB protocol (between honest Prover/Verifier) for a  secret $\mathbf{s}$ and error-parameter $\epsilon$. In other words, it is the distribution followed by $\langle A, \mathbf{z}=f(\mathbf{s}A) + \mathbf{v} \rangle$, where $A$ is a random matrix picked from $\{0,1\}^{k \times n}$ and $\mathbf{v}$ is a Bernoulli noise vector of length $D$. 
\item $U_{kn+D}$ represents uniformly distributed $(kn+D)$-length bitstrings. In other words, a bitstring $S$ from $U_{kn+D}$ satisfies $\mathrm{Pr}[S=\mathbf{g}] = 2^{-(kn+D)}\,;\, \forall \mathbf{g} \in \{0,1\}^{kn+D}$.
\item As already seen in Theorem \ref{lpntounld}, Algorithm $X$ denotes a UNLD solver algorithm. 
\item Algorithm $Z$ is an algorithm that is capable of forging the NLHB protocol in the passive model. Given $q$ bitstrings from the distribution ${\cal A}_{\mathbf{s},\epsilon,f}$, and a challenge matrix $A_1$, $Z$ can give a corresponding response $\mathbf{z_1}$ that will generate an ``Accept" response from the NLHB Verifier with non-negligible success probability. In other words, $d(\mathbf{z_1},f(\mathbf{s}A_1)) \leq u = \epsilon'D$  with non-negligible probability. 
\item The Advantage of $Z$ (denoted $Adv_Z^{NLHB-Attack}(k,\epsilon,u,f)$) is defined as the difference between probability of success of $Z$ and the probability of success of an attacker who gives random responses to the Verifier. Since the latter probability is $P_{FA}$, the probability of false-accept, and is negligible for large $D$, the advantage of $Z$ is almost the same as its probability of success. The advantage is a function of protocol parameters $k,\epsilon$ and $u$.
\item We also define an intermediate algorithm $Y$ for the purpose of the proof. The algorithm $Y$ is a distinguisher algorithm that can successfully distinguish between the distributions ${\cal A}_{\mathbf{s},\epsilon,f}$ (for secret $\mathbf{s}$) and $U_{kn+D}$. Given $q$ bitstrings from either distribution (which one, is unknown to $Y$), the algorithm $Y$ processes the bits in some way and outputs 0/1. The probability that $Y$ outputs 1 when the input is ${\cal A}_{\mathbf{s},\epsilon,f}$ (for a random $\mathbf{s}$) and the probability that $Y$ outputs 1 when the input is $U_{kn+D}$ differ significantly. That is,  
\begin{equation*}
\left| Pr\left[\mathbf{s} \leftarrow \lbrace 0,1 \rbrace ^{k} : Y^{{\cal A}_{\mathbf{s},\epsilon , f}}=1 \right]-Pr\left[Y^{U_{kn+D}}=1\right] \right| \geq \delta
\end{equation*}
for some non-negligible probability $\delta$. This difference in probabilities of $Y$ outputting 1 for the different distributions can be used to distinguish these two distributions. In the above equation, $Y^{U_{kn+D}}$ implies that the algorithm $Y$ is inputted bits that follow the distribution $U_{kn+D}$. The notation $Y^{{\cal A}_{\mathbf{s},\epsilon , f}}$ has a similar meaning.

Note that the difference between the probabilities above is for a random key $\mathbf{s}$. In other words, it is an average over all possible keys. 

\end{enumerate}
The reduction is in two steps. 
\begin{enumerate}
\item In the first step (given in Theorem \ref{thm2}), we prove a reduction from the UNLD problem to the problem of distinguishing between the distributions $U_{kn+D}$ and ${\cal A}_{\mathbf{s},\epsilon,f}$, i.e we construct $X$ using $Y$.
\item In the second step (Theorem \ref{thm3}), we provide a reduction from the problem of distinguishing between these distributions to the problem of forging the NLHB protocol, i.e, we construct $Y$ using $Z$.
\end{enumerate}
Thus, by using $Y$ as an intermediate, we prove a reduction from the UNLD problem to forging of NLHB protocol.

\subsubsection{Theorem \ref{thm2} Proof Outline} The algorithm $X$ uses $Y$ as follows to solve the UNLD problem. It first estimates $p$, the probability that $Y$ outputs 1 when given access to $U_{kn+D}$. To do this, it generates $q$ instances of $(kn+D)$ random bits and passes them to $Y$ (thus simulating $U_{kn+D}$ to $Y$) and obtains $Y$'s binary response. By repeating this $N$ times (for reasonably large $N$) and finding the fraction of 1s in the output, $X$ gets an estimate for $p$. Next, $X$ takes a bitstring $\langle A,\mathbf{z} \rangle$ from ${\cal A}_{\mathbf{s},\epsilon,f}$, to which it has access, by definition. Suppose $X$ wants to find $s_i$, the $i^{th}$ bit of the secret $\mathbf{s}$. $X$ adds a random vector $\mathbf{c}$ to the $i^{th}$ row of $A$. Let us call the resulting matrix $A'$. Also, let $hyb_i$ denote the distribution followed by the bitstring $\langle A',\mathbf{z} \rangle$. $X$ passes $q$ different instances of $hyb_i$ (for a given $i$) to $Y$ and obtains its binary output. Like before, it repeats this process $N$ times and estimates $p_i$, the probability that $Y$ outputs 1 when its input is $hyb_i$.

\begin{figure}[!htbp]
\begin{center}
\includegraphics[scale=1.2]{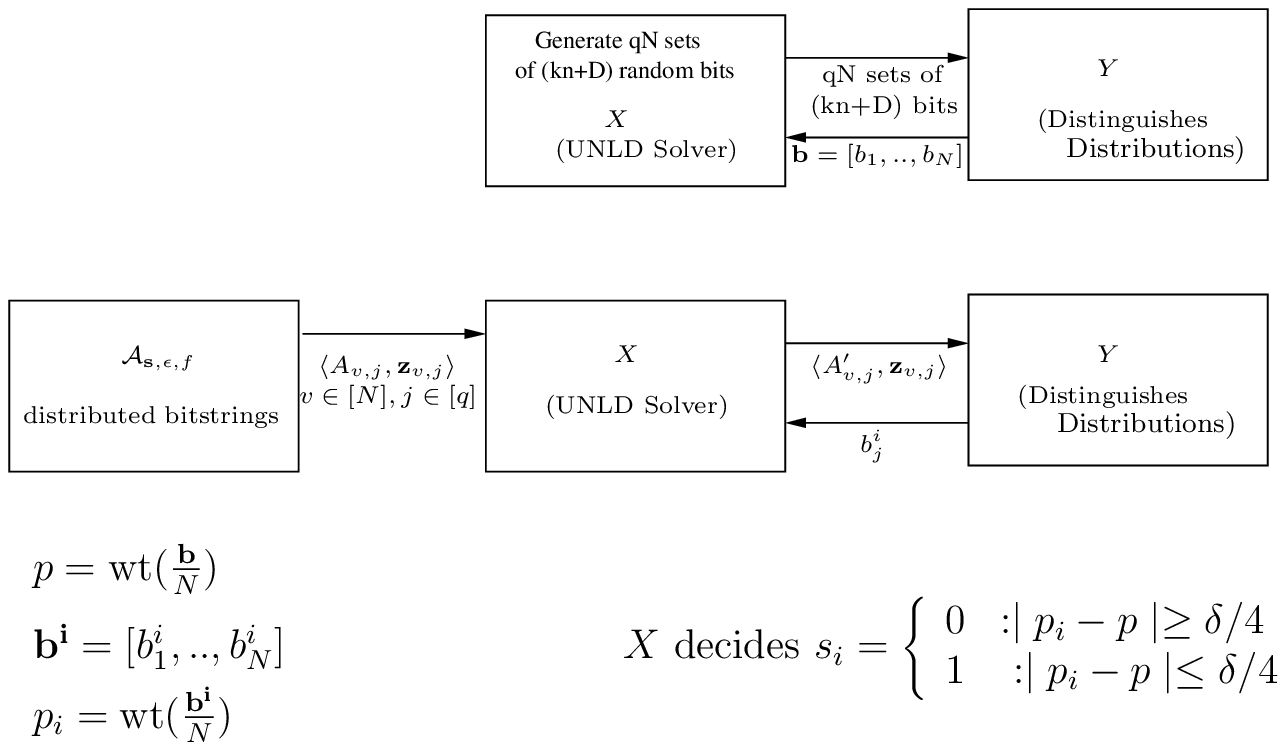}
\end{center}
\caption{Passing of Strings in the Proof of Theorem \ref{thm2}}
\label{thm2fig}
\end{figure}

If $s_i=0$, it is easy to see that $hyb_i={\cal A}_{\mathbf{s},\epsilon,f}$. Further, we prove in Lemma \ref{lemma1} below that if $s_i =1 $, then $hyb_i=U_{kn+D}$.  Since, by definition, $Y$ outputs 1 with significantly different probabilities for $U_{kn+D}$ and ${\cal A}_{\mathbf{s},\epsilon,f}$, $p_i$ will be very close to $p$ if $s_i=1$ (meaning $hyb_i=U_{kn+D}$) and away from $p$ if $s_i=0$. So, by estimating the probability of $Y$ outputting 1 and comparing it with $p$, $X$ can deduce $s_i$. By repeating this procedure for all $i \in [k]$, $X$ can solve the UNLD problem. We have shown this process in Figure \ref{thm2fig}. The subscripts of the passed values in the figure denote the $qN$ different instances being sent. We have omitted these subscripts in the above outline for the sake of readability. A more formal treatment is given in the Appendix.
\begin{lemma}[$hyb_i = U_{kn+D}$ if $s_i=1$]
Let $A$ be a randomly chosen $k \times n$ matrix. Let $\mathbf{s}$ be a random $k$-bit binary secret vector. Further, assign $\mathbf{z}=f(\mathbf{s}A) + \mathbf{v}$, where the bits of $\mathbf{v}$ are \textit{i.i.d} Bernoulli distributed. Now, let $\mathbf{c}$ be a randomly chosen (independent of all other factors) $n$-bit binary vector. For an arbitrary $i \in[k]$, let $A'$ denote the matrix formed by modifying only the $i^{th}$ row of $A$ as $(A)_i = (A)_i + \mathbf{c}$. If $hyb_i$ denotes the distribution of the bit-string $\langle A',\mathbf{z} \rangle$, then $hyb_i = U_{kn+D}$ if $s_i=1$.
\label{lemma1}
\end{lemma}
\begin{proof}
Consider the conditional probability $\mathrm{Pr}[\mathbf{z} = \mathbf{r} \mid A' = \hat{A}]$ for some $\hat{A}$ and an arbitrary $\mathbf{r} \in \{0,1\}^D$.
\begin{eqnarray*}
&&\mathrm{Pr}[\mathbf{z} = \mathbf{r} \mid A' = \hat{A}] = \mathrm{Pr}[f(\mathbf{s}A) + \mathbf{v}= \mathbf{r} \mid A' = \hat{A}] \\
&=& \mathrm{Pr}[f(\mathbf{s}A' + \mathbf{c}) + \mathbf{v}= \mathbf{r} \mid A' = \hat{A}] \quad \quad(\mathrm{Since} \, s_i=1)\\
&=& \mathrm{Pr}[f(\mathbf{s}\hat{A} + \mathbf{c}) + \mathbf{v}= \mathbf{r}]  
\end{eqnarray*}
We see that since $\mathbf{c}$ is chosen at random, independent of the other variables, $\mathbf{s}\hat{A} + \mathbf{c}$ varies uniformly in $\{0,1\}^{n}$. Consequently, $f(\mathbf{s}\hat{A} + \mathbf{c})$ varies uniformly at random in $\{0,1\}^D$ by Property 3 of $f$. So, we have
\begin{equation}
\mathrm{Pr}[\mathbf{z} = \mathbf{r} \mid A' = \hat{A}] = \mathrm{Pr}[f(\mathbf{s}\hat{A} + \mathbf{c}) + \mathbf{v}= \mathbf{r}]  = 2^{-D}
\label{p1}
\end{equation}
Further,
\begin{eqnarray*}
&&\mathrm{Pr}[\mathbf{z} = \mathbf{r}] = \sum_{\mathbf{x}}\mathrm{Pr}[\mathbf{z} = \mathbf{r} \mid \mathbf{v}=\mathbf{x}]\mathrm{Pr}[\mathbf{v}=\mathbf{x}] \\
&=& \sum_{\mathbf{x}}\mathrm{Pr}[f(\mathbf{s}A) = \mathbf{r} + \mathbf{x}]\mathrm{Pr}[\mathbf{v}=\mathbf{x}]
\end{eqnarray*}
Since $A$ is chosen at random and due to Property 3 of $f$, we have $\mathrm{Pr}[f(\mathbf{s}A) = \mathbf{r}+\mathbf{x}] = 2^{-D}$. So, 
\begin{equation}
\mathrm{Pr}[\mathbf{z} = \mathbf{r}] = \sum_{\mathbf{x}}2^{-D}\mathrm{Pr}[\mathbf{v}=\mathbf{x}] = 2^{-D}
\label{p2}
\end{equation}
From \eqref{p1} and \eqref{p2}, we see that $\mathbf{z}$ is independent of $A'$. 
So $\mathrm{Pr}[A' = \hat{A},\mathbf{z} = \mathbf{r}] = \mathrm{Pr}[A' = \hat{A}]\mathrm{Pr}[\mathbf{z} = \mathbf{r}] = 2^{-(kn+D)}$. Since this holds for any arbitrary $\mathbf{r} \in \{0,1\}^D$, $hyb_i = U_{kn+D}$ if $s_i=1$.
\end{proof}
Since the function $f$ plays an important role in this lemma, we have presented it here. Since the remaining proof of Theorem \ref{thm2} is not dependent on the function $f$, and the proof is adapted from \cite{katzandsmith,katzandshin}, we merely state the theorem here. Please refer to Appendix for detailed proofs.
\begin{theorem}\textit{(Reducing UNLD to Distinguishing ${\cal A}_{\mathbf{s},\epsilon,f}$ and $U_{kn+D}$):}
Suppose there exists a probabilistic polynomial-time  algorithm $Y$ taking $q$ bitstrings of an unknown distribution (either ${\cal A}_{\mathbf{s},\epsilon,f}$ or $U_{kn+D}$) and outputting $0/1$, running in time $t$, such that the probability of outputting 1 when its input is drawn from $U_{kn+D}$ and when its input is drawn from ${\cal A}_{\mathbf{s},\epsilon , f}$ differ by at least $\delta$, i.e 
\begin{equation*}
\left| Pr\left[\mathbf{s} \leftarrow \lbrace 0,1 \rbrace ^{k} : Y^{{\cal A}_{\mathbf{s},\epsilon ,f}}=1 \right]-Pr\left[Y^{U_{kn+D}}=1\right] \right| \geq \delta.
\end{equation*}
Then there exists $X$ taking $q'= O(q.\delta^{-2}log(k))$ bitstrings of ${\cal A}_{\mathbf{s},\epsilon,f}$ running in time $t'= O(t.k.\delta^{-2}log(k))$ such that
\begin{equation*}
Pr \left[ \mathbf{s} \leftarrow \lbrace 0,1 \rbrace ^{k} : X^{{\cal A}_{\mathbf{s},\epsilon ,f}} = \mathbf{s} \right] \geq \delta /4.
\end{equation*}
\label{thm2}
\end{theorem}
\begin{proof} Please refer Appendix A. \end{proof}
\subsubsection{Theorem \ref{thm3} Proof Outline} Having proven the hardness of of the distinguisher problem, we will now reduce it to the passive attack problem, thus proving the hardness of the passive attack problem. The distinguisher algorithm $Y$ can be constructed using algorithm $Z$ as follows. The algorithm $Y$ takes $q$ bitstrings from its unknown distribution, and passes them to $Z$. In each bitstring, $Z$ treats the first $kn$ bits to be the challenge matrix, and the next $D$ bits to be the corresponding response. This completes the query phase of $Z$. Now, $Y$ takes one last string $\langle \hat{A}, \mathbf{\hat{z}}\rangle$ from the unknown distribution. It passes $\hat{A}$, the first $kn$ bits of the string, to $Z$ as a challenge. Let $\mathbf{z'}$ be the response that $Z$ gives $Y$ to this challenge. Then, it can be shown that if the input distribution had been $U_{kn+D}$, then it is very unlikely that $\mathbf{z'}$ and $\mathbf{\hat{z}}$ are near each other in terms of Hamming distance. On the other hand, if the distribution had been ${\cal A}_{\mathbf{s},\epsilon,f}$, then these two are very likely to have Hamming distance below a threshold because of certain properties of the distribution of $Z$. So, if $Y$ outputs 1 whenever the Hamming distance $d(\mathbf{z'},\mathbf{\hat{z}})$ falls within an appropriately set threshold, the probability of $Y$ outputting 1 for the two distributions will vary significantly, thus fulfilling its requirements. This process is shown in Figure \ref{thm3fig}. Again, we state only the formal theorem here, and give the complete proof in the Appendix. Using this proof, and the reduction from algorithms $X$ to $Y$ in Theorem \ref{thm2}, we have a reduction from the UNLD problem to the passive attack problem. So, we conclude that the passive attack problem is hard.
\begin{figure}[!htbp]
\begin{center}
\includegraphics[scale=1.25]{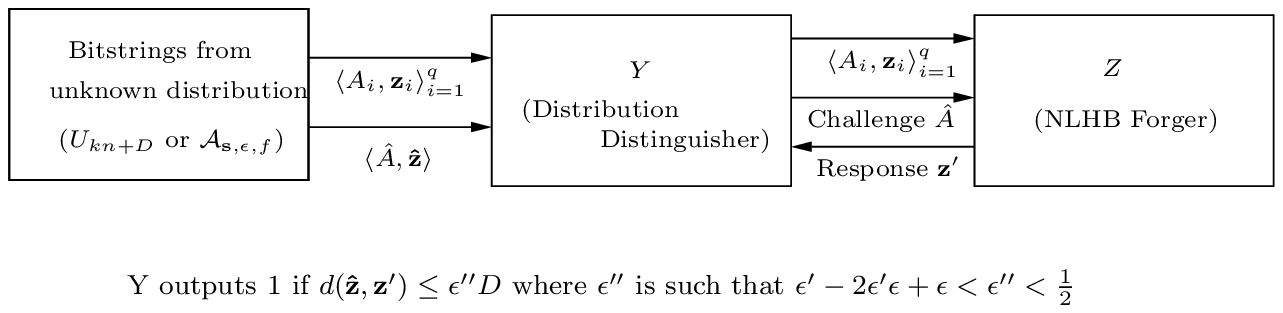}
\end{center}
\caption{Passing of Strings in the Proof of Theorem \ref{thm3}.}
\label{thm3fig}
\end{figure}
\begin{theorem}\textit{(Reduction From Distinguishing ${\cal A}_{\mathbf{s},\epsilon,f}$ and $U_{kn+D}$ To Forging NLHB Protocol in Passive Model):}
If $Adv^{NLHB-attack}_{Z}(k, \epsilon, u, f) = \delta$ is non-negligible for some polynomial time adversary $Z$, then the UNLD problem can be efficiently solved.
\label{thm3}
\end{theorem}
\begin{proof} Refer Appendix A.\end{proof}
\section{Implementation and Efficiency}
\label{imploff}
In this section, we consider the specific low-cost candidate for $f$ given in \eqref{fcand} and demonstrate how existing passive attacks on the HB protocol fail against the NLHB protocol. Then, we compare the Prover complexity of NLHB and HB protocols and demonstrate that the NLHB Prover is required to carry out lesser operations when compared to a HB prover that achieves the same level of security.
\subsection{Resistance Against Current Passive Attacks}
\label{lf2attackrepulsed}
Using the specific $f$ in \eqref{fcand}, we will show how the existing LF2 attack on LPN is ineffective on the NLHB protocol. Let $\mathbf{x}=[x_1,...,x_n] = \mathbf{s}A = [\mathbf{s}.\mathbf{a_1},...,\mathbf{s}.\mathbf{a_n}]$, where $[\mathbf{a_1},...,\mathbf{a_n}]$ are columns of $A$. Let $\mathbf{y}=f(\mathbf{x})$. Then, the passive adversary to NLHB has access to $\mathbf{z}= \mathbf{y} + \mathbf{v}$.

As explained in Section~\ref{lf2explained}, the LF2 (or BKW) algorithm works by repeatedly adding the columns of the matrix $A$ and obtaining the response corresponding to this new matrix by adding the responses corresponding to the added columns. We examine the result when the attacker does one column addition. Let the attacker modify $A$ into $A' = [\mathbf{a_1},...,\mathbf{a_j}+\mathbf{a_k},...,\mathbf{a_n}]$, i.e, he adds the $k^{th}$ column to the $j^{th}$ column. The corresponding matrix product between $\mathbf{s}$ and $A'$ will be $\mathbf{\overline{x}} = [x_1,x_2,...,x_j+x_k,...,x_n]$, i.e $\mathbf{\overline{x}}$ has the same bits as $\mathbf{x}$ except at the $j^{th}$ position, where it is $x_j + x_k$. Let $\mathbf{\overline{y}} = f(\mathbf{\overline{x}})$. Now let us compare the relation between the unnoised responses $\mathbf{y}$ and $\mathbf{\overline{y}}$. As can be seen, the only output bits getting affected by the change of matrix are the ones with indices $(j-3), (j-2), (j-1), j$. We readily see the following relationships.
\begin{eqnarray*}
y_{j-3}&=&x_{j-3} + x_{j-2}x_{j-1} + x_{j-1}x_j + x_jx_{j-2}.\\
y_{j-2}&=&x_{j-2} + x_{j-1}x_j + x_jx_{j+1} + x_{j+1}x_{j-1}.\\
y_{j-1}&=&x_{j-1} + x_{j}x_{j+1} + x_{j+1}x_{j+2} + x_{j+2}x_{j}\\
y_{j}&=&x_{j} + x_{j+1}x_{j+2} + x_{j+2}x_{j+3} + x_{j+3}x_{j+1}.\\
\overline{y}_{j-3}&=&x_{j-3} + x_{j-2}x_{j-1} + x_{j-1}(x_j+x_k) + (x_j+x_k)x_{j-2}.\\
\overline{y}_{j-2}&=&x_{j-2} + x_{j-1}(x_j +x_k) + (x_j+x_k)x_{j+1} + x_{j+1}x_{j-1}.\\
\overline{y}_{j-1}&=&x_{j-1} + (x_j+x_k)x_{j+1} + x_{j+1}x_{j+2} + x_{j+2}(x_j+x_k).\\
\overline{y}_{j}&=&x_{j} + x_k + x_{j+1}x_{j+2} + x_{j+2}x_{j+3} + x_{j+3}x_{j+1}.
\end{eqnarray*}
Let us denote the errors between these corresponding bits as $E_{j-3}, E_{j-2}, E_{j-1}, E_j$. From the above equations, we get 
\begin{eqnarray*}
E_{j-3} &=& y_{j-3}+\overline{y}_{j-3} = x_{j-1}x_k + x_kx_{j-2},\\
E_{j-2} &=& x_{j-1}x_k + x_kx_{j+1},\\
E_{j-1} &=& x_{j+1}x_k + x_kx_{j+2},\\
E_{j} &=& x_k.
\end{eqnarray*}
Each error term above is an unknown bit to the attacker, since he does not have access to either a noised or un-noised version of these terms. So, the attacker has to guess the error bits $E_{j-3},E_{j-2},E_{j-1},E_{j}$ that need to be added to the new response to get the estimate of responses corresponding to the new matrix. The amount of uncertainty involved in guessing these bits can be found from the entropy of $[E_{j-3},E_{j-2},E_{j-1},E_j]$. Since the bits $x_i$ are uniformly distributed, it can easily be seen that this entropy is equal to $2.5$ bits. So each time a column is added, the attacker has to guess 2.5 bits on an average. Since there are many such additions needed in the LF2 attack, this attack is no longer feasible against the NLHB protocol. 
In Table \ref{optimizef}, we give the values of the entropy of the bit-wise error terms for different choices of $p=2,3,4$ and functions. For $p=4$, we have shown only few functions out of the many that achieve the maximum entropy of 3.  As we can see, the entropy increases with increase in $p$, meaning that LF2 attacks are harder for higher $p$.
\begin{table}[h]
\centering
\begin{tabular}{|c | l | c |}
\hline
$\,\,p\,\,$ & Function Achieving Maximum entropy for given $p$ & Maximum Entropy Achieved for given $p$\\ \hline
2 & $y_i = x_i + x_{i+1}x_{i+2}$ & 2 \\ \hline
3 & $y_i = x_i + x_{i+1}x_{i+2} + x_{i+1}x_{i+3}$ & 2.5 \\ \hline
& $y_i = x_i + x_{i+1}x_{i+3} + x_{i+2}x_{i+3}$ & 2.5 \\ \hline
& $y_i = x_i + x_{i+1}x_{i+2} + x_{i+2}x_{i+3} + x_{i+3}x_{i+1}$ & 2.5 \\ \hline
4 & $y_i = x_i + x_{i+1}x_{i+4} + x_{i+2}x_{i+3}$ & 3 \\ \hline
 & $y_i = x_i + x_{i+1}x_{i+4} + x_{i+2}x_{i+4} + x_{i+3}x_{i+4}$ & 3 \\ \hline
  & $y_i = x_i + x_{i+1}x_{i+4} + x_{i+2}x_{i+3} + x_{i+3}x_{i+4}$ & 3 \\ \hline
\end{tabular}
\caption{Maximum Entropy Achieved Over All Functions For A Given $p$ and The Function Achieving This Maximum}.
\label{optimizef}
\end{table}

Similar arguments can be given for the infeasibility of the Imai~\cite{imai} attack, that also relies heavily on linearity. The Imai attack attempts to isolate bits of the response vector that are noise-free and process them to obtain the secret key through Gaussian elimination. However, due to the nonlinear nature of $f$, Gaussian elimination is not possible with NLHB. Instead, the attacker must solve around a system of ($k+ \gamma$) nonlinear equations in $k$ variables (one for each bit of the secret key). Considering that the number of variables is large, it would be interesting to see if such an attack that involves repeatedly solving systems of nonlinear equations can be efficiently mounted.

The infeasibility of passive attacks on the related HB protocol indicates that the NLHB protocol can achieve 80-bit security using keysizes smaller than 512 bits, which is the number of key bits needed by the HB protocol. Added to this, the fact that no passive solutions exist to the problem of decoding the random non-linear codes described here (and for decoding of random non-linear codes in general) implies that it is reasonable to use keysizes very close to 80 bits with this protocol. However, as a safe value for the key-size, we suggest using 128-bit keys as secrets to resist all known passive attacks on the HB protocol.
\subsection{Comparison of Prover Complexity of NLHB and HB}
Since each scalar multiplication in the binary field requires one AND gate and one binary addition requires one XOR gate, we calculate the Prover (or Verifier) algorithm's complexity in terms of binary additions and scalar multiplications. Further, since the complexity involved in adding noise is the same in both protocols, we compare the complexity involved in the calculation of the un-noised responses in the Prover (or Verifier). 

The response calculated by the HB protocol for a given random matrix challenge $A_{k \times n}$ is given by $\mathbf{z}=\mathbf{s}A + \mathbf{v}$. The matrix product $\mathbf{s}A$ requires $kn$ scalar multiplications and $(k-1)n$ (binary) additions for its calculation. Assuming that $\epsilon=.25$ and $\epsilon'=.348$, the length of the final vector to which noise is added should be $n=1164$ \cite{fouque}. The value of $k$ for the HB protocol to achieve 80-bit security is around $k=512$.

In the NLHB protocol, we have a $k \times (D+p)$ challenge-matrix $A$ which we use to find $\mathbf{s}A$. This requires $k(D+p)$ scalar multiplications and $(k-1)(D+p)$ additions. Further, for the NLHB protocol, we have to evaluate the function $f$ over this vector. If we assume that we use the function $f$ in (\ref{fcand}) (with $p=3$), we require $3D$ scalar multiplications and $3D$ additions for evaluating the function $f$. So to calculate $f(\mathbf{s}A)$, we need $k(D+3) + 3D = kD + 3k + 3D$ multiplications and $3D + (k-1)(D+3) = kD + 2D + 3k -3$ additions. Since we add a length-$D$ noise vector in NLHB, $D$ has to be 1164.

\begin{table}[h]
\centering
\begin{tabular}{|c|c|c|c|c|c|c|c|}
\hline
& $k$ & $\epsilon$ & $\epsilon'$ &Size of Challenge Matrix &Length Of Prover Response& Scalar Multiplications & Scalar Additions\\
\hline
HB & 512 & .25 & .348 &$ 512 \times 1164$&$n$=1164& 595968 & 594804 \\
\hline
NLHB & 128 & .25 & .348 & $128 \times 1167$&$D$=1164& 152868 & 151701 \\
\hline
\end{tabular}
\label{table1}
\caption{Comparison of Prover/Verifier Complexities between NLHB and HB for $f$ with $p=3$, False-Reject Probability $P_{FR} =2^{-40} $ and False-Accept Probability $P_{FA}=2^{-80}$ and 80-bit security.}
\end{table}

For the sake of comparing complexities, if we assume a high-security version of NLHB which uses  $k=512$, then we see that NLHB needs a total of 600996 multiplications and 599829 additions, whereas HB protocol requires 595968 multiplications and 594804 additions. This approximately represents a 0.85\% increase in the both the number of multiplications and additions. This shows us that even in comparison to a HB protocol using the same keysize as the NLHB protocol, the addition in complexity due to the introduction of $f$ is very small.

However, with $k=128$, the computation of noise-free NLHB response requires 152868 scalar multiplications and 151701 additions, which is far less than the number of computations needed for a HB protocol Prover to achieve the same level of security, which requires about 512 secret bits. 
\section{NLHB$^+$ Protocol : Extending NLHB To Achieve Security in ``DET" Model}
\hspace*{9mm}Though the NLHB protocol is secure against a passive adversary, it is not secure against an active attacker. An efficient active attack similar to the one demonstrated against HB can also be mounted on the NLHB protocol. So, in the spirit of the HB$^{+}$ protocol, we propose the NLHB$^{+}$ protocol to provide complete security in the DET model.

\begin{figure}[!htbp]
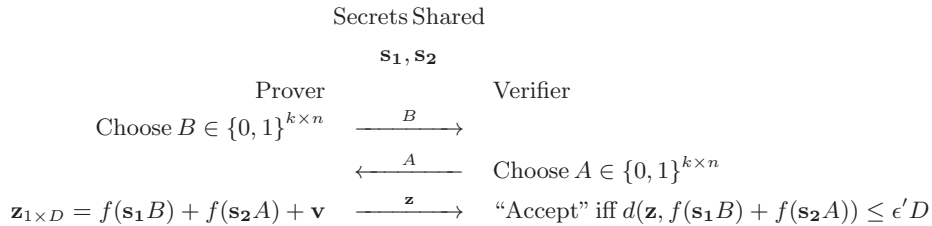

\begin{eqnarray*}
& \mathrm{Secrets} \, \mathrm{Shared} & \\
& \mathbf{s_{1}},\mathbf{s_{2}} &\\
\mbox{Prover} && \mbox{Verifier} \\
\mathrm{Choose} \, B\in \{ 0,1 \} ^{k \times n} & \stackrel{B}{\medrarrow} &\\
 &\stackrel{A}{\medlarrow}&   \mathrm{Choose} \, A \in \{ 0,1 \} ^{k \times n} \\
\mathbf{z}_{1 \times D}= f(\mathbf{s_{1}}B) +f(\mathbf{s_{2}}A) + \mathbf{v} & \stackrel{\mathbf{z}}{\medrarrow}& \mathrm{``Accept"} \, \mathrm{iff} \, d(\mathbf{z},f(\mathbf{s_{1}}B)+ f(\mathbf{s_{2}}A)) \leq \epsilon' D
\end{eqnarray*}
\caption{Parallelized version of the NLHB$^{+}$ protocol}
\label{figNLHB+}
\end{figure}

Figure \ref{figNLHB+} shows the NLHB$^{+}$ protocol. 
The Prover and Verifier share two secrets $\mathbf{s_{1}}$ and $\mathbf{s_{2}}$. Here, the authentication session is started when the Prover transmits a random $k \times n$ blinding matrix $B$ to the Verifier, which responds with a random $k \times n$ challenge matrix $A$. The Prover responds with $\mathbf{z}=f(\mathbf{s_{1}}B) + f(\mathbf{s_{2}}A) + \mathbf{v}$, where $f$ and $\mathbf{v}$ are as defined in the NLHB protocol. The Verifier replies with ``Accept" iff $d(\mathbf{z},f(\mathbf{s_{1}}B) + f(\mathbf{s_{2}}A)) \leq \epsilon'D$. NLHB$^+$ depends on the hardness of the UNLD problem for security against passive attacks. In addition, NLHB$^+$ is secure against active attacks in the ``DET" model as shown in the next section. 
\subsection{Security Proof for NLHB$^+$ In the ``DET" Model}
The security proof for NLHB$^+$ in the ``DET" model is given in Theorem \ref{formalthm4}, which gives a reduction to active attacks on the NLHB$^+$ protocol from the problem of differentiating ${\cal A}_{\mathbf{s},\epsilon,f}$ and $U_{kn+D}$. Since the latter problem has already been proven hard, this proves the hardness of active attacks. The strategy for Theorem~\ref{formalthm4} is broadly based on the proofs given in \cite{katzandsmith} with appropriate modifications to accomodate the function $f$. So, we simply give an outline of the theorem and the theorem statement here. Please refer the appendix for the complete proof.

First, we give some relevant definitions.
\begin{enumerate}
\item The algorithm $Z_+$ is a polynomial-time NLHB$^+$ active adversary. It is a two-phased algorithm. In its query phase, it takes a $k \times n$ random matrix $B$ as input. It then responds with a challenge matrix $A$ (which can be non-random). It should then be given the response that would be given by a legitimate NLHB$^+$ Prover for this $B$ and $A$, i.e, $\mathbf{z}=f(\mathbf{s_1}B) + f(\mathbf{s_2}A) + \mathbf{v}$ for secrets $\mathbf{s_1}$ and $\mathbf{s_2}$. In its challenge phase, $Z_+$ first sends a random blinding matrix $\hat{B}$ to the NLHB$^+$ verifier. It then receives a challenge matrix $\hat{A}$ from the Verifier and generates response $\mathbf{\hat{z}}$ that can generate ''Accept" from the NLHB$^+$ Verifier with non-negligible probability.
\item $Adv^{\mathrm{NLHB}^{+} attack}_{Z_{+}}(k, \epsilon, u, f)$ denotes the advantage for an active adversary $Z_{+}$ to the NLHB$^{+}$ protocol. It is defined as the difference in probabilities of success of $Z_+$ and a random attacker. Since the latter is $P_{FA}$, the probability of false-accept, and is negligible for large $D$, the advantage is almost the same as the probability of success of $Z_+$. The advantage is a function of the parameters $k$, $\epsilon$ and $u$.
\end{enumerate}
\subsubsection{Outline of Proof of Theorem \ref{formalthm4}}
The goal is to construct an algorithm $Y$ that can differentiate $U_{kn+D}$ from ${\cal A}_{\mathbf{s_1},\epsilon,f}$, which is the NLHB distribution with secret $\mathbf{s_1}$. To simulate a NLHB$^+$ Prover with two secrets to the algorithm $Z_+$, $Y$ generates a random vector $\mathbf{s_2}$ to be used as the second NLHB$^+$ secret. Now, $Y$ obtains the $(kn+D)$-length bitstring from the unknown distribution. We denote the first $kn$ bits of this string as $\overline{B}$ and the last $D$ bits as $\mathbf{\overline{z}}$. $Y$ passes $\overline{B}$ to $Z_+$, which responds with a challenge matrix $A$. Now, $Y$ responds with $\mathbf{z} = \mathbf{\overline{z}} + f(\mathbf{s_2}A)$. Note that if the input distribution had been ${\cal A}_{\mathbf{s_1},\epsilon,f}$, this is exactly the response expected by $Z_+$ for the secret pair ($\mathbf{s_1},\mathbf{s_2}$). This process if repeated $q$ times to complete the query phase of $Z_+$.

\begin{figure}[!htbp]
\begin{center}
\includegraphics[scale=.9]{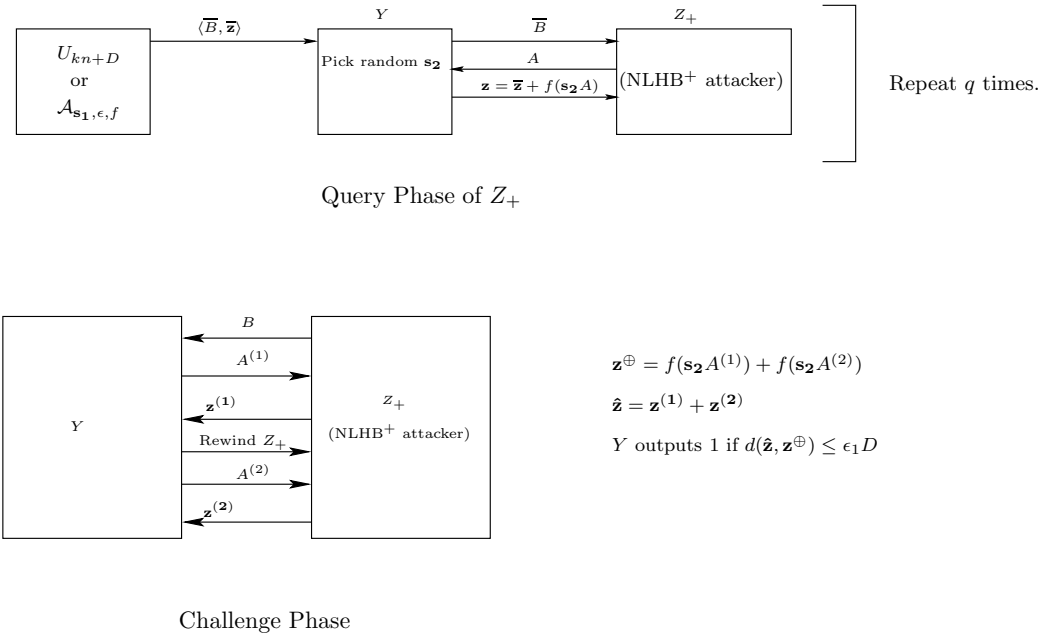}
\end{center}
\caption{Passing of Strings in Theorem \ref{formalthm4} Proof}
\label{thm4fig}
\end{figure}

In the challenge phase, the main trick used by $Y$ is that of rewinding $Z_+$. After receiving a blinding matrix $B$ from $Z_+$, it sends a matrix $A^{(1)}$ and receives response $\mathbf{z^{(1)}}$ from $Z_+$. Now, it rewinds $Z_+$ to the point where it sent $B$, and sends another challenge $A^{(2)}$ and receives $\mathbf{z^{(2)}}$ for the same $B$. By summing these responses $\mathbf{z^{(1)}}$ and $\mathbf{z^{(2)}}$, the effect of the unknown $\mathbf{s_1}$ can be removed. This is because $\mathbf{z^{\oplus}} = \mathbf{z^{(1)}}+\mathbf{z^{(2)}}$ is simply a noisy version of $\mathbf{\hat{z}} = f(\mathbf{s_2}A^{(1)}) + f(\mathbf{s_2}A^{(2)})$. Now it is easy to make statements about the distance between these two vectors. It can be shown that in case the distribution is $U_{kn+D}$, the probability that these vectors are ``close" (within a threshold distance) is low, and that if the distribution is ${\cal A}_{\mathbf{s},\epsilon,f}$, then this probability is a non-negligible function of $\delta$ (the advantage of $Z_+$), which is assumed to be non-negligible. So, $Y$ is able to output 1 with very different probabilities for the two distributions, thus helping us differentiate them. The passing of strings in this algorithm construction is shown in Figure \ref{thm4fig}. Since we already know that UNLD reduces to the problem of differentiating these distributions, we can now say that UNLD reduces to the active attack problem. So, the active attack problem is hard.
We now state the formal theorem and give the complete proof in the appendices.
\begin{theorem} If for some polynomial-time adversary $Z_{+}$, $Adv^{\mathrm{NLHB}^{+}attack}_{Z_{+}}(k, \epsilon, u, f)$ is non-negligible, then the UNLD problem can be efficiently solved.
\label{formalthm4}
\end{theorem}
\begin{proof} Refer Appendix B.\end{proof}
\section{Conclusion And Future Work}
In this paper, we have proven the hardness of a non-linear decoding problem that we call the UNLD problem and proposed the NLHB and NLHB$^+$ authentication protocols, which are variants of the HB and HB$^+$. These new protocols have better passive attack security than the HB and HB$^{+}$ protocols. They have a low-complexity and are most suited for RFID tags and other low-cost devices deployed in scenarios with attack monitoring.

In the future, it would be interesting to see if the MIM attacks~\cite{GRS,ouafi} (part of a prevention-based attack model) on the HB family of protocols can be prevented by making appropriate changes to the NLHB protocol. This will give rise to a protocol that can be used in systems where the presence of attacks are not monitored. Another useful line of exploration would be to study if the NLHB protocol offers any advantage compared to other protocols in real channels, since noise is an intrinsic part of the protocol's design.
\bibliographystyle{IEEETran}
\bibliography{references}

\appendix
\section{Formal Security Proof For NLHB In Passive Model}
\textbf{Theorem \ref{thm2} : Reduction From UNLD Problem To Distinguishing ${\cal A}_{\mathbf{s},\epsilon,f}$ and $U_{kn+D}$}\\
\newline
\textit{Suppose there exists a probabilistic polynomial-time  algorithm $Y$ taking $q$ bitstrings of an unknown distribution (either ${\cal A}_{\mathbf{s},\epsilon,f}$ or $U_{kn+D}$) and outputting $0/1$, running in time $t$, such that the probability of outputting 1 when its input is drawn from $U_{kn+D}$ and when its input is drawn from ${\cal A}_{\mathbf{s},\epsilon , f}$ differ by at least $\delta$, i.e 
\begin{equation}
\left| Pr\left[\mathbf{s} \leftarrow \lbrace 0,1 \rbrace ^{k} : Y^{{\cal A}_{\mathbf{s},\epsilon , f}}=1 \right]-Pr\left[Y^{U_{kn+D}}=1\right] \right| \geq \delta
\label{thmeq}
\end{equation}
Then there exists $X$ taking $q'= O(q.\delta^{-2}log(k))$ bitstrings of ${\cal A}_{\mathbf{s},\epsilon , f}$ running in time $t'= O(t.k.\delta^{-2}log(k))$ such that
\begin{equation*}
Pr \left[ \mathbf{s} \leftarrow \lbrace 0,1 \rbrace ^{k} : X^{{\cal A}_{\mathbf{s},\epsilon , f}} = \mathbf{s} \right] \geq \delta /4
\end{equation*}}
\textbf{Algorithm $X$ does the following:}
\begin{enumerate}
\item Pick $N=O(\delta^{-2}log(k))$.
\item $X$ chooses $w$ coins for $Y$ and uses these for the rest of the execution. \footnote{The coins act as the source of randomness in $Y$. In other words, the choosing of this coins can be thought of as $Y$ following one set of probabilistic decisions in its functioning out of the many possibilities.}
\item $X$ runs $Y^{U_{kn+D}}(w)$ $N$ times to obtain a bit string $\mathbf{b} = [b_{1},...,b_{N}]$. Let $p = \frac{\mathrm{wt}(\mathbf{b})}{N}$ be an estimate for the probability that $Y^{U_{kn+D}}$ outputs 1.
\item $X$ obtains $qN$ samples $\langle A_{v,j},\mathbf{z}_{v,j} \rangle, j=1,..,q\, ; \, v=1,..,N$ of distribution ${\cal A}_{\mathbf{s},\epsilon,f}$. ($q$ samples per response bit from $Y$ multiplied by the required $N$ responses required from $Y$). For $i \in [k]$:
\begin{enumerate}
\item Pick a random $n$-bit vector $\mathbf{c}_{v,j}$ for $j=1,..,q; \, v=1,..,N$. Modify $(A_{v,j})_{i}$, the $i^{th}$ row of $A_{v,j}$, as $(A_{v,j})_{i}= (A_{v,j})_{i}+\mathbf{c}_{v,j}$ to get a modified matrix $A'_{v,j}$. Pass the modified instance $\langle A'_{v,j},\mathbf{z}_{v,j} \rangle$ to $Y$ for $v=1,j=1,...,q$ to obtain its response $b^i_1$. Repeat this for $v=1,...,N$ to get the bit-string $\mathbf{b^i} = [b^i_1,...,b^i_N]$. Let $p_{i} = \frac{\mathrm{wt}(\mathbf{b^i})}{N}$ be an estimate of the probability that $Y$ returns a 1 when the $i^{th}$ row of $A_{v,j}; v \in [N],j \in [q]$ are modified.
\item If $\left| p_{i}-p \right| \geq \delta/4$ set $s'_{i}=0$, else set $s'_{i}=1$.
\end{enumerate}	
\item Output $\mathbf{s'}=(s'_{1},...,s'_{k})$.
\end{enumerate}
\textbf{Note 1: }There are three sources of randomness here. One is the randomness in the unknown key $\mathbf{s}$. The second is the randomness present in the decisions made by the existential algorithm $Y$. This randomness is denoted by the $w$ coins chosen for $Y$ at the beginning of the above algorithm. Both the key $\mathbf{s}$ and the $w$ coins are picked and held constant over the whole run of the algorithm. The third source of randomness comes from the picking of bitstrings from the distribution itself. 

\textbf{Note 2: }The difference in the probabilities of $Y$ outputting 1 for the two distributions in \eqref{thmeq}, is an averaged quantity. It is averaged over the key $\mathbf{s}$ and also on the randomness in the algorithm $Y$ itself. However, one run of the above algorithm only uses one instance of $\mathbf{s}$ and $w$. So, instead of using these averaged probabilities in our analysis, we should use the probabilities associated with the particular key and randomness $w$ used with this run of the algorithm $Y$. We will refer to the probability of $Y$ outputting 1 when it uses these particular $w$ coins as $\mathrm{Pr}[Y^{U_{kn+D}}(w) = 1]$ (Similarly $\mathrm{Pr}[Y^{{\cal A}_{\mathbf{s},\epsilon,f}}(w) = 1]$), i.e we use the argument $w$ to denote a particular set of decisions followed by $Y$.\\ 
\newline
\textbf{Analysis of the Algorithm:}

From the algorithm, we see that $p$ is an estimate for $\mathrm{Pr}[Y^{U_{kn+D}}(w) = 1]$. Further, if $hyb_i$ denotes the distribution of the $(kn+D)$ bits passed to $Y$ by $X$ in step 4(a), then $p_i$ is an estimate of $\mathrm{Pr}[Y^{hyb_i}(w) = 1]$. We now prove that for the chosen value of $N=O(\delta^{-2}log(k))$, $p$ and $p_i$ are very close estimates of these values. 

Consider $\mathrm{Pr}[\mid \mathrm{Pr}[Y^{U_{kn+D}}(w) = 1] - p \mid \leq \delta/16]$, i.e the probability of the event the actual value of $\mathrm{Pr}[Y^{U_{kn+D}}(w) = 1]$ and its estimate are within $\delta/16$ of each other. For ease of readability, let us denote $\mathrm{Pr}[Y^{U_{kn+D}}(w) = 1]$ by $Pr_U$. \\
\newline
\textbf{Accuracy of Estimates $p_i$ and $p$:} 

We know that $p = \frac{\mathrm{wt}(\mathbf{b})}{N}$. Each bit of $\mathbf{b}$ follows a Bernoulli distribution with mean $Pr_U$. So, $\mathrm{wt}(\mathbf{b})$ follows a Binomial distribution with mean $NPr_U$. So,
\begin{eqnarray*}
& \mathrm{Pr}[\mid p- Pr_U \mid \leq \delta/16] = \mathrm{Pr}[\mid \mathrm{wt}(\mathbf{b})- NPr_U \mid \leq N\delta/16],  \\
& = \mathrm{Pr}[NPr_U - N\delta/16 \leq \mathrm{wt}(\mathbf{b}) \leq NPr_U + N\delta/16], \\
& = 1 - \mathrm{Pr}[\mathrm{wt}(\mathbf{b}) > NPr_U + N\delta/16] - \mathrm{Pr}[\mathrm{wt}(\mathbf{b}) < NPr_U - N\delta/16].
\end{eqnarray*}
By applying Chernoff bounds on this Binomial random variable, we have
\begin{equation}
\mathrm{Pr}[\mid p- Pr_U \mid \leq \delta/16] \geq 1 - exp \left[ - \frac{N \delta^2}{768Pr_U} \right] - exp \left[ - \frac{N \delta^2}{512Pr_U} \right].
\label{eqmajvote}
\end{equation}
Now, we use $N = O(\delta^{-2}log(k))$. Let $d_1$ be a large constant such that $N \leq d_1\delta^{-2}log(k)$. Applying in \eqref{eqmajvote}, we get
\begin{equation}
\mathrm{Pr}[\mid p- Pr_U \mid \leq \delta/16] \geq 1 - \left(\frac{1}{k}\right)^{\frac{d_1}{768Pr_U}} - \left(\frac{1}{k}\right)^{\frac{d_1}{512Pr_U}}.
\label{eqq2}
\end{equation}
By similar reasoning, we also have, 
\begin{equation}
\mathrm{Pr}[\left| Pr_{hi} - p_{i} \right| \leq \delta /16] \geq 1 - \left(\frac{1}{k}\right)^{\frac{d_1}{768Pr_{hi}}} - \left(\frac{1}{k}\right)^{\frac{d_1}{512Pr_{hi}}},
\label{eqq3}
\end{equation}
where $Pr_{hi}$ is used to denote $Pr \left[ Y^{hyb_{i}}(w)=1 \right]$ for ease of readability.
We know that, for two independent events $E1$ and $E2$, if $\mathrm{Pr}[E1] \geq 1 - a$ and $\mathrm{Pr}[E2] \geq 1 - b$, then $\mathrm{Pr}[E1\cap E2] \geq 1 - a-b$. Applying this here, we see that $\mid p- Pr_U \mid \leq \delta/16$ and $\left| Pr_{hi} - p_{i} \right| \leq \delta /16$ (the latter for all $i$) hold simultaneously with probability 
\begin{equation*}
 \geq 1 - \left\{\left[\frac{1}{k}\right]^{\frac{d_1}{768Pr_U}} + \left[\frac{1}{k}\right]^{\frac{d_1}{512Pr_U}}\right\} - \sum_{i=1}^{k}\left\{\left[\frac{1}{k}\right]^{\frac{d_1}{768Pr_{hi}}} + \left[\frac{1}{k}\right]^{\frac{d_1}{512Pr_{hi}}}\right\}.
\end{equation*}

Let $l = \min \left\{\frac{1}{768Pr_u},\frac{1}{512Pr_u},\frac{1}{768Pr_{h1}},...,\frac{1}{768Pr_{hk}},\frac{1}{512Pr_{h1}},...,\frac{1}{512Pr_{hk}}\right\}$. Then the above expression can be lower-bounded as
\begin{equation*}
\geq 1 - (2k+2)\left( \frac{1}{k}\right)^{\frac{d_1}{l}} \geq 1 - 4k^{1 - \frac{d_1}{l}}
\end{equation*}
By choosing $d_1$ sufficiently large, ($d_1 = 4l$, say), we have that \eqref{eqq2} and \eqref{eqq3} hold simultaneously with probability $\geq \left( 1 - \frac{4}{k^3}\right) \geq \frac{1}{2}$ (for $k>1$).

In summary we have that the following equations hold simultaneously with probability at least $\frac{1}{2}$.
\begin{equation}
\mid p- \mathrm{Pr}[Y^{U_{kn+D}}(w)=1] \mid \leq \delta/16
\label{eq2}
\end{equation}
\begin{equation}
\mid p_i- \mathrm{Pr}[Y^{hyb_i}(w)=1] \mid \leq \delta/16, 1 \leq i \leq k.
\label{eq3}
\end{equation}
%
%
\newline
\textbf{Suppose $s_i=1$:}\\
\newline
Now, consider the case where $s_i = 1$. By Lemma \ref{lemma1}, in this case, $hyb_i = U_{kn+D}$. So, if both \eqref{eq2} and \eqref{eq3} hold, then for the case of $s_i=1$, we have
\begin{equation}
\mid p_i - p \mid \leq 2\delta/16 = \delta/8.
\label{eq6}
\end{equation}
\newline
\textbf{Suppose $s_i=0$:}\\
\newline
Now consider $s_i=0$. Then, since the $i^{th}$ row of $A'_{v,j}$ never plays a role in the output, $hyb_i = {\cal A}_{\mathbf{s},\epsilon, f}$. Now let us bound $\mid p_i - p \mid$ in this case.

From the definition of $Y$, we have
\begin{equation*}
\left| Pr\left[\mathbf{s} \leftarrow \lbrace 0,1 \rbrace ^{k} : Y^{{\cal A}_{\mathbf{s},\epsilon , f}}=1 \right]-Pr\left[Y^{U_{kn+D}}=1\right] \right| \geq \delta.
\end{equation*}
This is a bound on the difference between the probabilities on an average. Using a standard averaging argument, we now derive a bound for the difference between the probabilities of $Y$ outputting 1 in each case for the given instance of $\mathbf{s}$ and $w$.

\begin{lemma}By a standard averaging argument, with probability $\geq \delta/2$ over the choice of $\mathbf{s}$ and the random coins $w$, the following equation holds,
\begin{equation}
\label{eq1}
\left| Pr \left[ Y^{{\cal A}_{\mathbf{s},\epsilon, f}}(w)=1 \right] - Pr \left[ Y^{U_{kn+D}}(w)=1 \right] \right| \geq \delta/2,
\end{equation}
where the probabilities inside the equation are over the randomness involved in picking bitstrings from the distributions.
\label{lemma2}
\end{lemma}
\begin{proof} We prove this Lemma at the end of this Theorem. \end{proof}
Since we know that when $s_i=0$, $hyb_i = {\cal A}_{\mathbf{s},\epsilon,f}$, it follows that $\mathrm{Pr}[Y^{hyb_i}(w)=1] = \mathrm{Pr}[Y^{{\cal A}_{\mathbf{s},\epsilon,f}}(w)=1]$.
So, from \eqref{eq1}, we have,
\begin{equation}
\left| Pr \left[ Y^{hyb_i}(w)=1 \right] - Pr \left[ Y^{U_{kn+D}}(w)=1 \right] \right| \geq \delta/2.
\label{eq4}
\end{equation}
Rewriting \eqref{eq4}, we have 
\begin{eqnarray*}
\delta /2 \leq \left| Pr \left[ Y^{hyb_{i}}(w)=1 \right] -p_{i}+p_{i} - \right.   \left. Pr \left[ Y^{U_{kn+D}}(w)=1] -p +p\right] \right|,
\end{eqnarray*}
\begin{eqnarray*}
 \leq \left| p_{i}-p \right| + \left| Pr \left[ Y^{hyb_{i}}(w)=1 \right] -p_{i} \right|   + \left| Pr \left[ Y^{U_{kn+D}}(w)=1 \right]-p \right| ,
\end{eqnarray*}
\begin{equation*}
 \leq  \left| p_{i}-p \right| + \delta/16 + \delta/16,
 \end{equation*}
 assuming \eqref{eq2} and \eqref{eq3} hold. Finally, this implies
 \begin{equation}
  \left| p_{i}-p \right| \geq \delta/2 - 2.\delta/16 = 3.\delta/8.
 \label{eq5}
 \end{equation}
 So, if $s_i=0$, then $\mid p_i - p \mid \geq 3\delta/8$.\
 
So, in Step 5 of the algorithm $X$, if $|p_{i}-p| \leq \frac{\delta}{4}=\frac{2 \delta}{8}$, the estimated message bit is 1, else it is 0. Since \eqref{eq1} holds with probability atleast $\frac{\delta}{2}$ and \eqref{eq2} and \eqref{eq3} hold with a further probability of .5, the probability that \eqref{eq6} and \eqref{eq5} hold is at least $\frac{\delta}{4}$. Hence, algorithm $X$ succeeds with probability at least $\delta/4$.\qed 

\textbf{Proof For Lemma \ref{lemma2} : }
Let $R$ be the set of all possibilities for the key $\mathbf{s}$ and the $w$ coins. Then, by our definition of $Y$, we have, 
\begin{equation}
\sum_{\mathbf{s},w \in R} P[\mathbf{s},w] \left| Pr[ Y^{A_{\mathbf{s},\epsilon,f}}(w)=1] - Pr[Y^{U_{kn+D}}(w)=1] \right| \geq \delta.
\label{eqlr}
\end{equation}
Let the subset $R' \subset R$ be the set such that $ \forall r' = (\mathbf{s'},w')\in R'$, we have
\begin{equation}
\left| Pr[ Y^{A_{\mathbf{s'},\epsilon, f}}(w')=1] - Pr[Y^{U_{kn+D}}(w')=1] \right| < \delta/2.
\label{eq7}
\end{equation}
Then, in contradiction to Lemma 2, assume that the probability that $r'$ is picked at random is atleast $1-\delta/2$. Assuming that all $r'$ are equally likely to be picked, this implies that $\sum_{(\mathbf{s'},w') \in R'} Pr(\mathbf{s'},w') \geq 1- \delta/2$ and consequently, $\sum_{\mathbf{s},w \in R\backslash R'} Pr(\mathbf{s},w) < \delta/2$.\\
\newline
Now, splitting the left-hand-side (LHS) of \eqref{eqlr} into summations over $R'$ and $R' \backslash R$, we have
\begin{eqnarray*}
\mathrm{LHS} = \sum_{\mathbf{s'},w' \in R'}Pr[\mathbf{s'},w'] \left| Pr[ Y^{A_{\mathbf{s'},\epsilon, f}}(w')=1]  - \right. \left. Pr[Y^{U_{kn+D}}(w')= 1] \right| \\
+ \sum_{\mathbf{s},w \in R \backslash R'} Pr[\mathbf{s},w] \left| Pr[ Y^{A_{\mathbf{s},\epsilon,f}}(w)=1] - Pr[Y^{U_{kn+D}}(w)=1] \right|.
\end{eqnarray*}
By \eqref{eq7} and the fact that $\left| Pr[ Y^{A_{\mathbf{s},\epsilon,f}}(w)=1] - Pr[Y^{U_{kn+D}}(w)=1] \right| <1$ (because it contains probability terms), we have
\begin{equation*}
\mathrm{LHS} < (\delta/2)\sum_{\mathbf{s'},w' \in R'}Pr[\mathbf{s'},w'] + \sum_{\mathbf{s},w \in R \backslash R'} Pr[\mathbf{s},w] (1),
\end{equation*}
\begin{equation*}
=(\delta/2)(1-\sum_{\mathbf{s},w \in R \backslash R'}Pr[\mathbf{s},w]) + \sum_{\mathbf{s},w \in R \backslash R'} Pr[\mathbf{s},w].
\end{equation*}
Since we had $\delta \leq$ our original LHS from \eqref{eqlr}, this implies
\begin{equation*}
\delta < (\delta/2) + (1-\delta/2)\sum_{\mathbf{s},w \in R \backslash R'} Pr[\mathbf{s},w],
\end{equation*}
\begin{equation*}
\Rightarrow \sum_{\mathbf{s},w \in R \backslash R'} Pr[\mathbf{s},w] > \delta/2,
\end{equation*}
which contradicts our initial assumption about the set $R'$. So, by contradiction, Lemma 2 is true.\qed 

\textbf{Theorem \ref{thm3}: Reduction From Distinguishing ${\cal A}_{\mathbf{s},\epsilon,f}$ and $U_{kn+D}$ To Forging NLHB Protocol in Passive Model.}\\
\newline
\textit{If $Adv^{NLHB-attack}_{Z}(k, \epsilon, u, f) = \delta$ is non-negligible for some polynomial time adversary $Z$, then the UNLD problem can be efficiently solved.}\\
\newline
\textbf{Algorithm for Theorem \ref{thm3}:}
Given access to $Z$ which takes $q$ bitstrings of ${\cal A}_{\mathbf{s},\epsilon,f}$ and runs in time $t$ and forges the NLHB protocol with a passive attack, we construct an algorithm $Y$ that takes $q+1$ bitstrings from ${\cal A}_{\mathbf{s},\epsilon,f}$, and can distinguish between strings drawn from $U_{kn+D}$ and ${\cal A}_{\mathbf{s},\epsilon,f}$. $Y$ works like this.
\begin{enumerate}
\item $Y$ has access to bitstrings from either ${\cal A}_{\mathbf{s},\epsilon,f}$ or $U_{kn+D}$.
\item $Y$ draws $q$ strings $\langle A_{i},\mathbf{z_{i}}\rangle_{i=0}^{q}$ from this distribution. This is passed on to $Z$.
\item Now $Y$  obtains another sample pair $\langle \hat{A},\mathbf{\hat{z}}\rangle$ from the distribution (the first $kn$ bits of the bitstring drawn will represent $\hat{A}$ in case of either input distribution) and challenges $Z$ with $\hat{A}$. Let the received response be $\mathbf{z'}$.
\item $Y$ outputs 1 if $\mathbf{\hat{z}}$ and $\mathbf{z'}$ differ by atmost $u' = \epsilon''D$, i.e if $d(\mathbf{z'},\mathbf{\hat{z}}) \leq u'$, where $\epsilon''$ is some constant such that $\epsilon' - 2 \epsilon \epsilon' + \epsilon < \epsilon'' < \frac{1}{2}$.
\end{enumerate}
\textbf{Analysis of the algorithm:} If $Y$'s input distribution is $U_{kn+D}$, the probability that $Y$ outputs 1 is $p_{U}(1)=\sum_{i=0}^{u'} \binom{D}{i} 2^{-D}$. Since $\epsilon'' < .5$, $p_{U}(1)$ is negligible if $D$ is large enough.\\
\newline
Let $\mathbf{z^{*}}=f(\mathbf{s}\hat{A})$. Let $\mathbf{w}$ and $\mathbf{e}$ be error vectors corresponding to $\mathbf{z'}$ and $\mathbf{\hat{z}}$, i.e $\mathbf{z'}=\mathbf{z^{*}} + \mathbf{w}$ and $\mathbf{\hat{z}}=\mathbf{z^{*}} + \mathbf{e}$. Then, $d(\mathbf{z',z^{*}}) \leq u$ implies that $\mathrm{wt}(\mathbf{w}) \leq u$ and $d(\mathbf{z',\hat{z}}) \leq u'$ implies that $\mathrm{wt}(\mathbf{w}+\mathbf{e}) \leq u'$.

Consider the conditional probability $\mathrm{Pr}[d(\mathbf{z',\hat{z}}) \leq u' \mid d(\mathbf{z',z^{*}}) \leq u] = \mathrm{Pr}[\mathrm{wt}(\mathbf{w}+\mathbf{e}) \leq u' \mid \mathrm{wt}(\mathbf{w}) \leq u]$.
It is possible to show that
\begin{equation}
\mathrm{Pr}[\mathrm{wt}(\mathbf{w}+\mathbf{e}) \leq u' \mid \mathrm{wt}(\mathbf{w}) \leq u] \geq \mathrm{Pr}[\mathrm{wt}(\mathbf{w}+\mathbf{e}) \leq u' \mid \mathrm{wt}(\mathbf{w}) = u]
\label{thm3eq1}
\end{equation}
We give a proof for \eqref{thm3eq1} at the end of this proof.
We will now consider the right-hand-side of \eqref{thm3eq1} and prove that it is negligibly close to 1.
The conditional expectation of $\mathrm{wt}(\mathbf{w}+\mathbf{e})$ given $\mathrm{wt}(\mathbf{w})=u$ is given by
\begin{eqnarray*}
E \left[ \mathrm{wt}(\mathbf{w}+\mathbf{e})\mid \mathrm{wt}(\mathbf{w})=u \right]= u.(1-\epsilon) + (D-u) \epsilon \\
= (\epsilon' - 2 \epsilon \epsilon' + \epsilon)D
\end{eqnarray*}
Since $\epsilon'' > \epsilon' - 2 \epsilon \epsilon' + \epsilon$, we see that the following Chernoff bound holds:
\begin{equation*}
Pr[\mathrm{wt}(\mathbf{w}+\mathbf{e}) >(1+ \Delta)\mu \mid \mathrm{wt}(\mathbf{w})=u] \leq \left(\frac{\mathrm{exp}(\mu \Delta)}{ (1+\Delta)^{(1+\Delta)\mu}}\right),
\end{equation*}
where $\mu = (\epsilon' - 2 \epsilon \epsilon' + \epsilon)D$ is the mean of the random variable $\mathrm{wt}(\mathbf{w}+\mathbf{e})$ given that $\mathrm{wt}(\mathbf{w})=u$, $(1+ \Delta)\mu = \epsilon''D$, which imply that $\Delta = \frac{\epsilon''}{\epsilon' - 2 \epsilon \epsilon' + \epsilon}-1$. 


So we have 
\begin{eqnarray*}
&&\mathrm{Pr}[\mathrm{wt}(\mathbf{w}+\mathbf{e}) \leq u' \mid \mathrm{wt}(\mathbf{w}) \leq u] \\
& \geq &\mathrm{Pr}[\mathrm{wt}(\mathbf{w}+\mathbf{e}) \leq u' \mid \mathrm{wt}(\mathbf{w}) = u] \geq \left[ 1- \left(\frac{\mathrm{exp}(\mu \Delta)}{ (1+\Delta)^{(1+\Delta)\mu}}\right)\right].
\end{eqnarray*}
We also know that $\mathrm{Pr}[\mathrm{wt}(\mathbf{w}+\mathbf{e}) \leq u'] =  \mathrm{Pr}[\mathrm{wt}(\mathbf{w}+\mathbf{e}) \leq u' \mid \mathrm{wt}(\mathbf{w}) \leq u]\mathrm{Pr}[\mathrm{wt}(\mathbf{w}) \leq u]$. By the definition of the NLHB forger $Z$, we know that  $\mathrm{Pr}[d(\mathbf{z'},\mathbf{z^{*}}) = \mathrm{wt}(\mathbf{w}) \leq u] \geq (\delta + P_{FA})$, where $P_{FA}$ denotes the probability of success of an attacker who responds with a random response ($P_{FA}$ is known to be negligibly small at high $D$).
So, we have 
\begin{equation}
\mathrm{Pr}[\mathrm{wt}(\mathbf{w}+\mathbf{e}) \leq u'] \geq (\delta + P_{FA})\left[ 1- \left(\frac{\mathrm{exp}(\mu \Delta)}{ (1+\Delta)^{(1+\Delta)\mu}}\right)\right].
\end{equation}
Consequently, the difference in the probabilities of $Y$ outputting 1 for the two distributions is at least 
\begin{equation}
(\delta + P_{FA})\left[ 1- \left(\frac{\mathrm{exp}(\mu \Delta)}{ (1+\Delta)^{(1+\Delta)\mu}}\right)\right] - \sum_{i=0}^{u'} \binom{D}{i} 2^{-D}.
\label{advdiff}
\end{equation}
Using suitable protocol parameters $D, \epsilon, \epsilon'$ (say, $D=1000, \epsilon=.25, \epsilon'=.348$ \cite{fouque}), we see that the value in \eqref{advdiff} is negligibly close to $\delta$. This proves that $Y$ can be constructed from $Z$.
\qed
\textbf{Proof For \eqref{thm3eq1}} :
We see that (by applying Bayes rule)
\begin{eqnarray*}
&& \mathrm{Pr}[\mathrm{wt}(\mathbf{w}+\mathbf{e}) \leq u' \mid \mathrm{wt}(\mathbf{w}) \leq u]  = \mathrm{Pr}[\mathrm{wt}(\mathbf{w}) \leq u \mid \mathrm{wt}(\mathbf{w}+\mathbf{e}) \leq u'] \frac{\mathrm{Pr}[\mathrm{wt}(\mathbf{w}+\mathbf{e}) \leq u']}{\mathrm{Pr}[\mathrm{wt}(\mathbf{w}) \leq u]}, \\
&& =\sum_{i=0}^{u} \mathrm{Pr}[\mathrm{wt}(\mathbf{w}) =i \mid \mathrm{wt}(\mathbf{w}+\mathbf{e}) \leq u'] \frac{\mathrm{Pr}[\mathrm{wt}(\mathbf{w}+\mathbf{e}) \leq u']}{\mathrm{Pr}[\mathrm{wt}(\mathbf{w}) \leq u]}.
\end{eqnarray*}
Applying Bayes Rule again, the above expression reduces to
\begin{equation*}
\mathrm{Pr}[\mathrm{wt}(\mathbf{w}+\mathbf{e}) \leq u' \mid \mathrm{wt}(\mathbf{w}) \leq u]= \sum_{i=0}^{u} \mathrm{Pr}[ \mathrm{wt}(\mathbf{w}+\mathbf{e}) \leq u' \mid \mathrm{wt}(\mathbf{w}) =i] \frac{\mathrm{Pr}[\mathrm{wt}(\mathbf{w}) =i]}{\mathrm{Pr}[\mathrm{wt}(\mathbf{w}) \leq u]}.
\end{equation*}
The random variable $\mathrm{wt}(\mathbf{w} + \mathbf{e}) \mid \mathrm{wt}(\mathbf{w})=i$ is the sum of the bits of $\mathrm{wt}(\mathbf{w}+\mathbf{e})$ and has a mean $\mu_i=(1- \epsilon)i + (D-i)\epsilon$. Since the bits of $(\mathbf{w}+\mathbf{e})$ are independent, $(\mathrm{wt}(\mathbf{w} + \mathbf{e}) \mid \mathrm{wt}(\mathbf{w})=i) \sim N(\mu_i,\sigma^2)$, where $\sigma^2 = D\epsilon(1-\epsilon)$. So, the probability $\mathrm{Pr}[ \mathrm{wt}(\mathbf{w}+\mathbf{e}) \leq u' \mid \mathrm{wt}(\mathbf{w}) =i]$ can be given by the Cumulative Distribution Function (CDF) $[1 - Q\left(\frac{u'- \mu_i}{\sigma}\right)]$ where the function $Q(.)$ is the tail-probability of $N(0,1)$ defined as $Q(\alpha) = \frac{1}{\sqrt{2 \pi}}\int_{\alpha}^{\infty}e^{-\frac{x^2}{2}}dx$. Since $Q$-function is a decreasing function, and $\mu_i > \mu_{i-1}$,  $\mathrm{Pr}[ \mathrm{wt}(\mathbf{w}+\mathbf{e}) \leq u' \mid \mathrm{wt}(\mathbf{w}) =i]$ is a decreasing function of $i$. So, we have 
\begin{equation*}
\mathrm{Pr}[\mathrm{wt}(\mathbf{w}+\mathbf{e}) \leq u' \mid \mathrm{wt}(\mathbf{w}) \leq u]  \geq \mathrm{Pr}[ \mathrm{wt}(\mathbf{w}+\mathbf{e}) \leq u' \mid \mathrm{wt}(\mathbf{w}) =u]\sum_{i=0}^{u}  \frac{\mathrm{Pr}[\mathrm{wt}(\mathbf{w}) =i]}{\mathrm{Pr}[\mathrm{wt}(\mathbf{w}) \leq u]}. 
\end{equation*}
This implies that 
\begin{equation}
\mathrm{Pr}[\mathrm{wt}(\mathbf{w}+\mathbf{e}) \leq u' \mid \mathrm{wt}(\mathbf{w}) \leq u] \geq \mathrm{Pr}[\mathrm{wt}(\mathbf{w}+\mathbf{e}) \leq u' \mid \mathrm{wt}(\mathbf{w}) = u].
\end{equation}
\qed
From Theorems \ref{lpntounld} to \ref{thm3}, we can see that, if UNLD is hard, then it is hard to forge a Prover of the NLHB protocol in polynomial-time, making NLHB computationally secure in the passive attack model. 

\section{Security Proof For NLHB$^{+}$ In DET Model:}
\textbf{Theorem \ref{formalthm4}: Reduction From UNLD Problem To Active Attack on $\mathrm{NLHB}^{+}$:} \textit{If for some polynomial-time adversary $Z_{+}$, $Adv^{\mathrm{NLHB}^{+}attack}_{Z_{+}}(k, \epsilon, u, f) = \delta$ is non-negligible, the UNLD problem can be efficiently solved.}\\
\newline	
To prove this, we show how to build the algorithm $Y$ that can differentiate between distributions $U_{nk+D}$ and $A_{\mathbf{s_{1}},\epsilon, f}$ (for secret $\mathbf{s_1}$) using access to a NLHB$^+$ adversary $Z_+$ inthe ``DET" model. This proof strategy is based on \cite{katzandsmith}.\\
\newline
\textbf{Algorithm for $Y$}:
\begin{enumerate}
\item $Y$ chooses $\mathbf{s}_{2}$ at random from $\{ 0,1 \}^{k}$. During the query phase of $Z_{+}$, $Y$ draws the bitstring $\langle \overline{B},\mathbf{\overline{z}} \rangle$ (as usual, irrespective of the input distribution, the first $kn$ bits will form $\overline{B}$) from its unknown input distribution ($U_{kn+D}$ or $A_{\mathbf{s_{1}},\epsilon, f}$) and passes $\overline{B}$ to $Z_{+}$. $Z_{+}$ replies with challenge $A$. In response, $Y$ sends $\mathbf{z}=\mathbf{\overline{z}}+f(\mathbf{s_{2}}A)$ to $Z_{+}$. This is repeated $q$ times.
\item In its challenge phase, $Z_{+}$ sends a matrix $B$ as blinding matrix to $Y$. $Y$ challenges $Z_{+}$ with random matrix $A^{(1)}$ and receives response $\mathbf{z^{(1)}}$ from $Z_{+}$. Now, $Y$ rewinds $Z_{+}$ and challenges it with another random matrix $A^{(2)}$ and receives $\mathbf{z^{(2)}}$ in response.
\item Let $\mathbf{z^{\oplus}}=\mathbf{z^{(1)}}+\mathbf{z^{(2)}}$. Further, let $\mathbf{\hat{z}}=f(\mathbf{s_{2}}A^{(1)})+f(\mathbf{s_{2}}A^{(2)})$. $Y$ outputs 1 if $\mathbf{z^{\oplus}}$ and $\mathbf{\hat{z}}$ differ in fewer than $u'=\epsilon_{1}D$ entries. ($\epsilon_{1}$ to be defined).
\end{enumerate}
\textbf{Analysis of the Algorithm:} When $Y$'s input is $U_{kn+D}$, $\mathbf{\overline{z}}$ is uniformly distributed. Hence $\mathbf{z}=\mathbf{\overline{z}}+f(\mathbf{s}_{2}A)$ is also uniformly distributed and independent of $\mathbf{s}_{2}$. So no information about $\mathbf{s_{2}}$ reaches $Z_{+}$ in the query phase. This means that, as far as $Z_{+}$ is concerned, $\mathbf{\hat{z}}$ is uniformly distributed in the random code $C = \{f(\mathbf{s_2}A^{(1)}) + f(\mathbf{s_2}A^{(2)})\}_{s_2}$. Now, we show that $\mathrm{Pr}[d(\mathbf{z^{\oplus}},\mathbf{\hat{z}}) \leq \epsilon_1D]$ is negligibly small for large $D$.

Consider the Hamming Ball $B$ of radius $\epsilon_1D$ centred at $\mathbf{z^{\oplus}}$. Let $X$ be the number of codewords of $C$ present in this Hamming Ball. When the matrices $A^{(1)}$ and $A^{(2)}$ are picked, they are picked uniformly at random. This means, because of Property 3 of $f$ (uniform inputs $\Rightarrow$ uniform outputs), the vectors in code $C$ form a random code. We now apply the Markov Inequality on $X$.
\begin{equation}
\mathrm{Pr}[X \geq p] \leq \frac{E(X)}{p}, 
\label{markov}
\end{equation}
where $E(X)$ is the mean of $X$.
Now consider $\mathrm{Pr}[d(\mathbf{z^{\oplus}},\mathbf{\hat{z}}) > \epsilon_1D]$. We see that
\begin{eqnarray}
\mathrm{Pr}[d(\mathbf{z^{\oplus}},\mathbf{\hat{z}}) > \epsilon_1D]  &=& 
 \mathrm{Pr}[d(\mathbf{z^{\oplus}},\mathbf{\hat{z}}) > \epsilon_1D \mid X < p]\mathrm{Pr}[X < p] + \mathrm{Pr}[d(\mathbf{z^{\oplus}},\mathbf{\hat{z}}) > \epsilon_1D \mid X \geq p]\mathrm{Pr}[X \geq p] \notag \\
&\geq & \mathrm{Pr}[d(\mathbf{z^{\oplus}},\mathbf{\hat{z}}) > \epsilon_1D \mid X < p]\mathrm{Pr}[X < p] \label{eq20}
\end{eqnarray}
Consider $\mathrm{Pr}[d(\mathbf{z^{\oplus}},\mathbf{\hat{z}}) > \epsilon_1D \mid X < p]$. This can be written as 
\begin{eqnarray*}
\mathrm{Pr}[d(\mathbf{z^{\oplus}},\mathbf{\hat{z}}) > \epsilon_1D \mid X < p] &=& \mathrm{Pr}[X<p \mid d(\mathbf{z^{\oplus}},\mathbf{\hat{z}}) > \epsilon_1D]\frac{\mathrm{Pr}[d(\mathbf{z^{\oplus}},\mathbf{\hat{z}}) > \epsilon_1D]}{\mathrm{Pr}[X<p]}, \\
&=& \sum_{i=0}^{p-1} \mathrm{Pr}[X=i \mid d(\mathbf{z^{\oplus}},\mathbf{\hat{z}}) > \epsilon_1D]\frac{\mathrm{Pr}[d(\mathbf{z^{\oplus}},\mathbf{\hat{z}}) > \epsilon_1D]}{\mathrm{Pr}[X<p]},\\
&=&\sum_{i=0}^{p-1} \mathrm{Pr}[d(\mathbf{z^{\oplus}},\mathbf{\hat{z}}) > \epsilon_1D \mid X=i]\frac{\mathrm{Pr}[X=i]}{\mathrm{Pr}[X<p]}.
\end{eqnarray*}
Notice that the quantity $\mathrm{Pr}[d(\mathbf{z^{\oplus}},\mathbf{\hat{z}}) > \epsilon_1D \mid X=i]$ will decrease with increase in $i$. This is because, with more codewords of $C$ within the Hamming ball $B$, the higher is the chance that $\mathbf{\hat{z}}$ lies within the Hamming Ball $B$, and so, higher is the chance that the distance between $\mathbf{\hat{z}}$ and $\mathbf{z^{\oplus}}$ is within $\epsilon_1D$. So, we can write 
\begin{eqnarray*}
\mathrm{Pr}[d(\mathbf{z^{\oplus}},\mathbf{\hat{z}}) > \epsilon_1D \mid X < p] &\geq & \sum_{i=0}^{p-1} \mathrm{Pr}[d(\mathbf{z^{\oplus}},\mathbf{\hat{z}}) > \epsilon_1D \mid X=p]\frac{\mathrm{Pr}[X=i]}{\mathrm{Pr}[X<p]}, \\
&=& \mathrm{Pr}[d(\mathbf{z^{\oplus}},\mathbf{\hat{z}}) > \epsilon_1D \mid X=p].
\end{eqnarray*}
 So, we have from \eqref{eq20} that
\begin{eqnarray}
\mathrm{Pr}[d(\mathbf{z^{\oplus}},\mathbf{\hat{z}}) > \epsilon_1D]  &\geq & \mathrm{Pr}[d(\mathbf{z^{\oplus}},\mathbf{\hat{z}}) > \epsilon_1D \mid X<p]\mathrm{Pr}[X < p], \notag \\ &\geq & \mathrm{Pr}[d(\mathbf{z^{\oplus}},\mathbf{\hat{z}}) > \epsilon_1D \mid X=p]\mathrm{Pr}[X < p], \notag \\
 &=& \mathrm{Pr}[\mathbf{\hat{z}} \not\in B \mid X = p]\mathrm{Pr}[X < p]. \label{fivex}
\end{eqnarray}
We know from the Markov inequality in \eqref{markov}, that $\mathrm{Pr}[X < p]$ is lower bounded by $\left( 1- \frac{E(X)}{p}\right)$. The mean number of codewords from $C$, which are part of the Hamming Ball $B$ is given by $E(X) = \left( \frac{\mid B \mid}{2^D}\right)2^k$. So \eqref{fivex} becomes 
\begin{equation}
\mathrm{Pr}[d(\mathbf{z^{\oplus}},\mathbf{\hat{z}}) > \epsilon_1D] \geq \left( 1- \frac{\mid B \mid 2^{k-D}}{p}\right).
\label{fivey}
\end{equation}
Out of the $2^k$ codewords of $C$, the probability that $\mathbf{\hat{z}}$ is one of the $p$ codewords in $B$ is given by $\frac{p}{2^k}$. So, the probability that $\mathbf{\hat{z}}$ does not belong to the Hamming ball $B$ when it is known that $B$ has exactly $p$ codewords of $C$, is given by $\left(1 - \frac{p}{2^k}\right)$. 
So, 
\begin{equation*}
\mathrm{Pr}[d(\mathbf{z^{\oplus}},\mathbf{\hat{z}}) > \epsilon_1D] \geq \left(1 - \frac{p}{2^k}\right)\left( 1- \frac{\mid B \mid 2^{k-D}}{p}\right).
\end{equation*}
Pick $p = 2^{3k/4}$, say.
Then
\begin{equation*}
\mathrm{Pr}[d(\mathbf{z^{\oplus}},\mathbf{\hat{z}}) > \epsilon_1D] \geq \left(1 - 2^{-k/4}\right)\left( 1- \mid B \mid 2^{\frac{k}{4}-D} \right).
\end{equation*}
We notice that $\mid B \mid$, the number of vectors in a Hamming Ball of radius $\epsilon_1D$ is given by $\mid B \mid = \sum_{i=0}^{\epsilon_1D} \binom{D}{i}$. So, 
\begin{equation*}
\mathrm{Pr}[d(\mathbf{z^{\oplus}},\mathbf{\hat{z}}) > \epsilon_1D] \geq \left(1 - 2^{-k/4}\right)\left( 1-  2^{\frac{k}{4}-D}\sum_{i=0}^{\epsilon_1D} \binom{D}{i} \right).
\end{equation*}
Since $\epsilon_1 < \frac{1}{2}$, this bound tends to 1 asymptotically with $D$. So, the probability $\mathrm{Pr}[d(\mathbf{z^{\oplus}},\mathbf{\hat{z}}) \leq \epsilon_1D]$ becomes negligibly small. So, in case the input distribution to $Y$ is $U_{kn+D}$, the probability of $Y$ outputting 1 is negligible.\\
\newline
When $Y$'s input distribution is $A_{\mathbf{s_{1}},\epsilon, f}$ for randomly chosen $\mathbf{s_{1}}$, $Y$ perfectly simulates the $\mathrm{NLHB}^{+}$ protocol to $Z_{+}$ during the query phase. Let $w$ denote the randomness involved in simulating the query phase of $Z_{+}$, which includes $Z_{+}$'s randomness, the randomness in choosing ($\mathbf{s_{1},s_{2}}$), and the randomness in responding to $Z_{+}$'s queries. Let $(\delta_{w} + P_{FA})$ be the probability that $Z_{+}$ successfully impersonates the Prover in second phase when the randomness is $w$. Then $Z_{+}$ correctly replies to both queries $A^{(1)}$ and $A^{(2)}$ with probability $(\delta_{w} + P_{FA})^{2}$. The overall probability that $Z_{+}$ successfully responds to both sets of queries is 
\begin{equation}
\mathrm{E}_{w}((\delta_{w}+ P_{FA})^{2}) \geq (\mathrm{E}_{w}(\delta_{w}+P_{FA}))^{2} = (\delta + P_{FA})^{2}
\end{equation}
using Jensen's inequality and $\mathrm{E}_w$ denotes expectation over $w$. Conditioned on this event, we show that for an appropriate $\epsilon_{1}$, $\mathbf{z^{\oplus}}$ and $\mathbf{\hat{z}}$ differ by fewer than $u'$ entries with a constant probability (proven below). So $Y$ outputs 1 with probability $\Omega ((\delta + P_{FA})^{2})$, which implies that $Y$ can distinguish $U_{kn+D}$ and ${\cal A}_{\mathbf{s_1},\epsilon,f}$ with non-negligible probability. This concludes the proof of Theorem \ref{formalthm4}. \qed
\textbf{Pf. for  $\mathbf{z^{\oplus}}$ and $\mathbf{\hat{z}}$ differing by $\leq u'$ entries:} Set $\frac{1}{2} > \epsilon_{1} > \frac{1}{2} (1-(1-2\epsilon')^{2})$. Fixing all randomness, let $f_{Z_{+}}$ denote the mapping that the adversary does from a challenge matrix $A$ to the response $\mathbf{z}$ in the second phase. Since we are looking at the process after $B$ has been fixed, $B$ is not an argument to the function $f_{Z_{+}}$. Let $f_{correct}$ denote $f(\mathbf{s_{1}}A) + f(\mathbf{s_{2}}B)$. Define $\Delta(A)=f_{Z_{+}}(A)+f_{correct}(A)$. We say that $A$ is a \textit{good} query matrix if $\mathrm{wt}(\Delta(A)) \leq u$, i.e if $Z_{+}$ successfully impersonates the Prover for that matrix. Let $D_{\Delta}$ denote the distribution of $\Delta (A)$ over all good query matrices. Note that by definition, for all $\Delta(A)$ in $D_{\Delta}$, $\mathrm{wt}(\Delta(A)) \leq u$.\\
\newline
Let $\Delta^{(1)} = \Delta(A^{(1)})$ and $\Delta^{(2)} = \Delta(A^{(2)})$. Then,
\begin{eqnarray*}
\Delta^{(1)} + \Delta^{(2)} \, =  \, f_{Z_{+}}(A^{(1)})+f_{Z_{+}}(A^{(2)}) +f_{correct}(A^{(1)})+f_{correct}(A^{(2)}).
\end{eqnarray*}
Using $f_{correct}(A^{(1)})+f_{correct}(A^{(2)})= f(\mathbf{s_{2}}A^{(1)})+f(\mathbf{s_{2}}A^{(2)}) \, = \, \mathbf{\hat{z}}$
and $f_{Z_{+}}(A^{(1)})+ f_{Z_{+}}(A^{(2)}) = \mathbf{z^{\oplus}}$, we see that $d(\Delta^{(1)},\Delta^{(2)}) \leq u'$ whenever $d(\mathbf{z^{\oplus},\hat{z}}) \leq u'$. We now analyse the probability that $d(\Delta^{(1)},\Delta^{(2)}) \leq u'$.\\

Using arguments based on the Johnson bound as in \cite{katzandsmith}, we can show that $Pr[d(\Delta^{(1)},\Delta^{(2)}) \leq u'] > \frac{1}{c^{2}}$, where $c=\frac{1-\delta_{eps}}{\gamma^{2}-\delta_{eps}}+1$, $\delta_{eps} = 1-2\epsilon_{1}$ and $\gamma = 1- 2\epsilon'$.  So $Y$ outputs 1 with probability at least $\frac{1}{2c^2}(\delta + P_{FA})^2$ when the input distribution is ${\cal A}_{\mathbf{s_1},\epsilon,f}$. \qed
So, the difference in probabilities of $Y$ in the proff of Theorem \ref{formalthm4} outputting a 1 for the two distributions ${\cal A}_{\mathbf{s_{1}},\epsilon,f}$ and $U_{kn+D}$ is at least 
\begin{equation}
(\frac{1}{2c^{2}}(\delta+P_{FA})^{2})-2^{-k/4}-  2^{\frac{k}{4}-D}\sum_{i=0}^{\epsilon_1D} \binom{D}{i} + 2^{-D}\sum_{i=0}^{\epsilon_1D} \binom{D}{i}.
\end{equation}
We see that this difference in probabilities tends to the non-negligible quantity $\frac{1}{2c^{2}} \delta^{2}$ asymptotically with $D$ (and for fixed reasonably large $k$).\\
\newline
Thus Theorems \ref{thm2} and \ref{formalthm4} together show a reduction from the UNLD problem to the problem of active attack on NLHB$^{+}$ protocol. Since the UNLD problem is known to be hard now,  the active attack problem is also hard.
\section{Proof For Uniformity of Function $f$}
\label{proofofuniformity}
\textbf{Theorem :$f$ is a Balanced Function}: If the input to the function $f$ is uniformly distributed, so is its output.\\
\textbf{Proof}
We first prove that each bit of the output is balanced. For this, we consider $\mathrm{Pr}[y_i=1]$. 
\begin{eqnarray*}
\mathrm{Pr}[y_i=1]  & = & \mathrm{Pr}[ x_i + g(x_{i+1},...,x_{i+p}) = 1], \notag \\ & = & \frac{1}{2}\mathrm{Pr}\left[g(x_{i+1},...,x_{i+p})=1 \mid x_i=0\right] + \frac{1}{2}\mathrm{Pr}\left[g(x_{i+1},...,x_{i+p})=0 \mid x_i=1\right]. \notag 
\end{eqnarray*}
Since the input vector is uniform, the bits of $\mathbf{x}$ are independent. So, this is equal to
\begin{eqnarray}
&=& \frac{1}{2}\mathrm{Pr}\left[g(x_{i+1},...,x_{i+p})=1\right] + \frac{1}{2}\mathrm{Pr}\left[g(x_{i+1},...,x_{i+p})=0 \right], \notag \\
&=&\frac{1}{2}.
\label{balanced}
\end{eqnarray}
So each bit of the output is balanced. Now, we use this to prove our theorem. To this end, we first define the following vectors. Let $\mathbf{y^i} = [y_{D-i+1},..,y_D]$ be the vector containing the last $i$ bits of $\mathbf{y}$. So $\mathbf{y^D} = \mathbf{y}$. Let $\mathbf{a} =[a_1,...,a_D]$ be an arbitrary constant $D$-bit vector. We also define $\mathbf{a^i} = [a_{D-i+1},..,a_D]$ similar to $\mathbf{y^i}$. Now consider the probability $\mathrm{Pr}[\mathbf{y^i} = \mathbf{a^i}]$. 
\begin{eqnarray*}
\mathrm{Pr}[\mathbf{y^i} = \mathbf{a^i}] &=& \mathrm{Pr}[\mathbf{y^i} = \mathbf{a^i} \mid x_{D-i+1}=0]\mathrm{Pr}[x_{D-i+1}=0] \\ & & \quad \quad \quad \quad \quad +  \mathrm{Pr}[\mathbf{y^i} = \mathbf{a^i} \mid x_{D-i+1}=1]\mathrm{Pr}[x_{D-i+1}=1]. 
\end{eqnarray*}
Since the input is uniformly distributed, this is equal to
\begin{eqnarray}
&=& \frac{1}{2}\mathrm{Pr}[\mathbf{y^i} = \mathbf{a^i} \mid x_{D-i+1}=0] + \frac{1}{2}\mathrm{Pr}[\mathbf{y^i} = \mathbf{a^i} \mid x_{D-i+1}=1], \notag \\
 &=&\frac{1}{2}\mathrm{Pr}[g(x_{D-i+2},...,x_{D-i+p+1}) = a_{D-i+1}, \mathbf{y^{i-1}} =\mathbf{a^{i-1}} \mid x_{D-i+1}=0], \notag\\
 && \quad + \frac{1}{2}\mathrm{Pr}[g(x_{D-i+2},...,x_{D-i+p+1}) = a_{D-i+1}+1, \mathbf{y^{i-1}} =\mathbf{a^{i-1}} \mid x_{D-i+1}=1]. \notag\\
\end{eqnarray}
We point out that in the vector $\mathbf{y^i}$, only the bit $y_{D-i+1}$ is dependent on $x_{D-i+1}$. Since both \\$g(x_{D-i+2},...,x_{D-i+p+1})$ and $\mathbf{y^{i-1}}$ are independent of $x_{D-i+1}$, we can remove the conditioning from the above equation. So the above expression becomes,
\begin{eqnarray}
&& \frac{1}{2}\left(\mathrm{Pr}[g(x_{D-i+2},...,x_{D-i+p+1}) = a_{D-i+1}, \mathbf{y^{i-1}} =\mathbf{a^{i-1}}] \right.\notag \\
 && \qquad \qquad+ \mathrm{Pr}[g(x_{D-i+2},...,x_{D-i+p+1}) = a_{D-i+1}+1, \mathbf{y^{i-1}} =\mathbf{a^{i-1}}]\left.\right). \notag \\
\end{eqnarray}
Now $g(x_{D-i+2},...,x_{D-i+p+1})$ takes binary values 0 and 1. So, by summing the joint probability of \\$g(x_{D-i+2},...,x_{D-i+p+1})$ and $\mathbf{y^{i-1}}$ over these values, we are effectively finding the marginal probability of $\mathbf{y^{i-1}}$. So, from the expressions in Eqn. 2 and 3, we have 
\begin{equation}
\mathrm{Pr}[\mathbf{y^i}=\mathbf{a^i}] = \frac{1}{2} \left( \mathrm{Pr}[\mathbf{y^{i-1}} = \mathbf{a^{i-1}}]\right).
\end{equation}
Plugging $i = D$ in the above equation, and expanding, we have
\begin{eqnarray}
\mathrm{Pr}[\mathbf{y^D}=\mathbf{a^D}] &=& \frac{1}{2}\left(\mathrm{Pr}[\mathbf{y^{D-1}}=\mathbf{a^{D-1}}]\right)\notag \\
&=& \frac{1}{2^2}\left( \mathrm{Pr}[\mathbf{y^{D-2}}=\mathbf{a^{D-2}}] \right) \notag \\
&&\vdots \notag \\
&=&\frac{1}{2^{D-1}}\left(\mathrm{Pr}[\mathbf{y^1} = \mathbf{a^1}]\right) = \frac{1}{2^{D-1}}\left(\mathrm{Pr}[y_D = a_D]\right) \notag \\
&=& \frac{1}{2^D}
\end{eqnarray}
from \eqref{balanced}.
Since this proof holds for any $\mathbf{a^i}$, the output of $f$ is uniformly distributed. \qed
\section{Notations}
\begin{itemize}
\item All vectors are denoted in bold letters. Scalars are denoted in normal text.
\item $] 0,\frac{1}{2}[$ denotes open-interval from 0 to $\frac{1}{2}$.
\item $\{ 0,1 \} ^{x}$ denotes the space of all binary vectors of length $x$.
\item $\{ 0,1 \} ^{x \times y}$ denotes the space of all binary matrices of size $x \times y$.
\item $\mathrm{wt}(\mathbf{x})$ denotes the Hamming weight of the binary vector $\mathbf{x}$. This is equal to the number of non-zero entries in $\mathbf{x}$.
\item $d(\mathbf{x},\mathbf{y})$ denotes the Hamming distance between binary vectors $\mathbf{x}$ and $\mathbf{y}$. This is equal to the number of places where $\mathbf{x}$ and $\mathbf{y}$ differ.
\item $GF(2)$ denotes Galois Field with two entries.
\item In this paper, $+$ is used to denote XOR addition which is the addition over $GF(2)$.
\item $\binom{n}{k} = \frac{n!}{k! (n-k)!}$, where $n!$ denotes factorial.
\item $\mbox{U}{\leftarrow}$ denotes "picked uniformly at random from".
\item When a distribution is superscripted over an algorithm, (for e.g. $X^{A}$) this means that the algorithm $X$ has input following the distribution $A$.
\item ${\cal A}_{\mathbf{s},\epsilon,f}$ denotes the distribution followed by the $(kn+D)$-length bitstrings that form the transcript of one authentication session between honest NLHB prover and verifier, for a shared secret $\mathbf{s}$. 
\item $U_{kn+D}$ denotes the distribution of uniformly distributed $(kn+D)$-length bitstrings.
\item For a set $R$ and its subset $R' \subset R$, $R \backslash R'$ denotes the set containing all the elements in $R$ that are not in $R'$.
\end{itemize}
\end{document}